\documentclass[preprints,article,accept,moreauthors,pdftex]{Definitions/mdpi} 

 \usepackage{longtable}

\firstpage{1} 
\pubvolume{xx}
\issuenum{1}
\articlenumber{5}
\pubyear{2019}
\copyrightyear{2019}
\history{Received: date; Accepted: date; Published: date}

\Title{Game Theory in defence applications: a review}

\Author{Edwin Ho$^{1}$,  Arvind Rajagopalan$^{2}$,   Alex Skvortsov$^{2}$, Sanjeev Arulampalam$^{2}$,  and \\  Mahendra Piraveenan$^{1,*}$}

\AuthorNames{Mahendra Piraveenan}

\address{%
$^{1}$ \quad Faculty of Engineering, University of Sydney, Sydney,  NSW 2006, Australia \\
$^{2}$ \quad Maritime Division, Defence Science and
Technology (DST) Group, Australia \\}

\corres{Correspondence: mahendrarajah.piraveenan@sydney.edu.au}

\abstract{This paper presents a succinct review of attempts in the literature to use game theory to model decision making scenarios relevant to defence applications.  Game theory has been proven as a very effective tool in modelling decision making processes of intelligent agents, entities, and players. It has been used to model scenarios from diverse fields such as economics, evolutionary biology, and computer science. In defence applications, there is often a need to model and predict actions of hostile actors, and players who try to evade or out-smart each other.  Modelling how the actions of competitive players shape the decision making of each other is the forte of game theory. In past decades, there have been several studies which applied different branches of game theory to model a range of defence-related scenarios. This paper provides a structured review of such attempts, and classifies existing literature in terms of the kind of warfare modelled, the types of game used, and the players involved. in terms of the warfares modelled, we  recognise that most papers which apply game theory in defence settings are concerned with command and control warfare, and can be further classified into papers dealing with (i)  Resource Allocation Warfare (ii) Information Warfare (iii)  Weapons Control Warfare (iv) Adversary Monitoring Warfare. We also observe that most of the reviewed papers are concerned with sensing, tracking, and large sensor networks, and the studied problems have parallels in sensor network analysis in the civilian domain. In terms of the games used, we classify the reviewed papers into papers that use  non-cooperative or cooperative games,  simultaneous or sequential games, discrete or continuous games, and  non-zero sum or zero-sum games. Similarly, papers are also classified into two-player, three-player or multi-player game based papers. We also explore the nature of players and the construction of payoff functions in each scenario. Finally, we also identify gaps in literature where game theory could be fruitfully applied in scenarios hitherto unexplored using game theory. The presented analysis provides a concise summary about the  state-of-the-art with regards to the use of game theory in defence applications, and highlights the benefits and limitations of game theory in the considered scenarios.}


\begin{document}



\section{Introduction} \label{introduction}

Game Theory has become one of the conventional theoretical frameworks to model  important decision making processes in many aspects of our life. The well-known examples include economy, social sciences, finance, population dynamics and epidemics (see \citep{piraveenan2019applications, peldschus2008experience,schotter2008economic,morrow1994game,hammerstein1994game} and references therein).  Since the seminal work of John Von Neumann, John Nash, and others \citep{von1944game, nash1950equilibrium, nash1951non}, it has been well recognised that there is an optimal  strategy  in the context of complex interactions (games)   between two or more parties (players) that can lead to a predictable outcome (payoff). In practical situations this outcome can often be quantitative and amenable to arithmetic operations (cost,  number of infected people, number of vaccinated people etc),  but often it  can be qualitative in nature  (such as risk, readiness level, health state etc).

Application of  game theory in defence has a long and diversified history  ranging from  the design of real time military systems (e.g, applied in missile interception) to the support of strategic decisions  on large defence investments and acquisitions. There is  extensive literature on specific theoretical methods and tools and their defence applications \citep{martin1978selective,machiavelli1984prince, homer2005iliad, tzu2014art}. We believe that the review of this literature is of interest to the community dealing  with operational analysis and data-driven decision support.  This is the main motivation for the presented study.

Game theory \citep{myerson2013game,osborne2004introduction} elevates military strategies and decisions with a more holistic and value-laden analysis of the situation \citep{martin1978selective}.  For the military, the potential scenarios amenable to game theoretic analysis include the rapidly growing applications of  autonomous intelligent systems, and game theory provides a comprehensive mathematical framework that greatly enhances the capabilities of decision making of the people who use these systems. Because of its potential, research into game theory is burgeoning, with a few papers beginning to emerge in the literature for this military research niche. The aim of this review is to assist researchers in utilising the body of knowledge in game theory to develop smarter and safer software systems for defence
practitioners. Given that the state of such research is still in an incipient phase, we do this by drawing connections between existing military knowledge with the nascent possibilities that game theory offers so that it can become a more widely understood and considered framework in military software systems.

Although it is not overly extensive, the body of literature around game theory in the military has covered a notable portion of the different forms of engagement and combat. These papers cover past, present and future scenarios: from predictive strategies in potential hostile situations to analytical assessments in hindsight of military standoffs thousands of years ago. Game theory has demonstrated
the capacity to be useful in any such military scenario. However, rapid technological progression has led to consistently new frontiers of military engagement, each of these possessing its own complex systems. The overarching areas that have been addressed are Tracking systems (across all domains), Aerial combat, Ground combat, Homefront warfare, Cyber warfare and Space systems. Notably, applications of game theory in  Naval warfare have been few, and an exploration into the future research into areas like this will be discussed later in the review. Within each of these areas there are a myriad of possibilities for new and innovative systems: different agents, different weapons, different control structures - and each of these could be enriched with game theoretic analysis. While Haywood's and Thunholm's treatises on game theory used in military decision making provide coverage of several different game types \citep{haywood1954military, thunholm2005planning}, there does not seem to be a paper which addresses the use of game theory in the military across each of the respective fields in the new context of military systems built on complex and high performance processing and algorithms. Our aim is to present the literature in such a way that it addresses all of the functions of game theory in military software systems in each key domain.

This review highlights the scope and utility of each analysed paper by presenting it in terms of the essential game components: players, game types, strategies and the key parameters of their payoff functions. It will act as both an annotated bibliography  as well as a framework to understand and plan further research into the area. It will also layout the fundamental tenets that are considered by players in every military decision making  scenario, as well as how they impact the decisions that are made by military personnel and systems, either while competing with hostile players or while cooperating with friendly players. This will make it possible for most military scenarios to be viewed as  games, and  can provide, at the very least, an interesting new perspective on familiar  military situations.

The rest of the paper is structured as follows.  Section \ref{background} will discuss the basic defence principles which are elaborated by the papers that we review, as well as introduce basic concepts of game theory.  Section \ref{analysis}  investigates and analyses the literature and summarises the findings and associations in each of the papers.  Section \ref{classification} elaborates our multi-dimensional classification of the literature based on the observations made in the previous section. Section \ref{futurework} identifies the gaps in the literature, and based on this, highlights opportunities for future research in niche, particularly areas of defence research which could benefit from the application of game theory where game theory has not been often applied so far. Finally, section \ref{conclusions} summarises our findings and classifications, and provides  broad conclusions.

\section{Background} \label{background}

Ideologies, beliefs and knowledge about war have been shaping human knowledge and philosophy for centuries. The great works of Sun Tzu, Homer and Machiavelli \citep{machiavelli1984prince, homer2005iliad, tzu2014art} have not only established a foundation for knowledge etched into the essence of military decision making, but also provided insight into sociology and social psychology \citep{andrzejewski2013military}. The military forms a core power bloc for many civilisations, and is instrumental to both the growth of influence for existing nations, and the birth of new nations \citep{janowitz1964military}. The military deals with conflicts in real time, plans for the future, as well as reviews past engagements - and every single one of these activities has an impact on society \citep{degrasse2016military}. This review therefore by necessity addresses many facets of military conflict across multiple physical domains, and the major decisions which define each domain will be further clarified below. With respect to game theory for all of these domains,  the value of targets, the value of resources, and the priority of objectives are usually the key parameters that shape the payoff functions and strategies which in turn define the games that we use in modelling.

In this section, we discuss the concepts in defence science and technology, as well as game theory, which are necessary to understand and analyse the literature in the presented niche. First of all, let as consider the broad domains of defence and national security which are considered in this review. They can be summarised, as shown in Table \ref{classstructure}.

\begin{longtable}[c]{|c|ccccc|}
\caption{Classification System used in this review}
\label{classstructure}\\
\hline
Military's Main Method & \multicolumn{5}{c|}{Command and Control Warfare} \\ \hline
\endhead
Period & \multicolumn{3}{c|}{Traditional (T)} & \multicolumn{2}{c|}{Modern (M)} \\ \hline
Military Field & \multicolumn{1}{c|}{Land (L)} & \multicolumn{1}{c|}{Air (A)} & \multicolumn{1}{c|}{Sea (S)} & \multicolumn{1}{c|}{Cyber (C)} & Space (S) \\ \hline
Major Military Field & \multicolumn{5}{c|}{\begin{tabular}[c]{@{}c@{}}Resource Allocation Warfare (RAW) \\ or\\ Information Warfare (IW)\\ or\\ Weapon Control Warfare (WCW) \\ or \\ Adversary Monitoring Warfare (AMW)\end{tabular}} \\ \hline
\multicolumn{1}{|l|}{Game Theory Category 1} & \multicolumn{5}{c|}{\begin{tabular}[c]{@{}c@{}}Non-Cooperative (NCo)\\ or \\ Cooperative (Co)\end{tabular}} \\ \hline
\multicolumn{1}{|l|}{Game Theory Category 2} & \multicolumn{5}{c|}{\begin{tabular}[c]{@{}c@{}}Sequential (Seq)\\ or \\ Simultaneous (Sim)\end{tabular}} \\ \hline
\multicolumn{1}{|l|}{Game Theory Category 3} & \multicolumn{5}{c|}{\begin{tabular}[c]{@{}c@{}}Discrete (D)\\ or \\ Continuous (C)\end{tabular}} \\ \hline
Game Theory Category 4 & \multicolumn{5}{c|}{\begin{tabular}[c]{@{}c@{}}Zero Sum (ZS)\\ or \\ Non-Zero Sum (NZS)\end{tabular}} \\ \hline
Number of Players & \multicolumn{5}{c|}{\begin{tabular}[c]{@{}c@{}}2 player (2P) \\ or\\  3-player (3P) \\ or \\ N-player (NP)\end{tabular}} \\ \hline
\end{longtable}

As shown in the table \ref{classstructure}, in this review, the focus is primarily on `command and control' warfare, where decision making is  critical. However, command and control warfare has applicability in traditional  modes of warfare, such as land, sea, and air warfare, as well as modern modes of warfare, such as space and cyber warfare. At an orthogonal level, command and control warfare could also be sub-divided into Resource Allocation Warfare (RAW), Information Warfare (IW), Weapons Control Warfare (WCW), as well as Adversary Monitoring Warfare (AMW). Since these concepts are extensively used in our classification of literature, let us briefly introduce them first.

\subsection{Warfare types}

\subsubsection{Land warfare}

Technology is a dictating force of warfare, and technology is not as imperative to land warfare as it is to other domains\citep{roland1991technology}. The technology that has impacted land warfare has been static, and avoids the exposure of human resources if possible \citep{paxson1979interactions, edmonds1992land}. Interpersonal combat at a physical level is much less prevalent nowadays, making way for a greater focus on positioning strategy. The literature which applied game theory to ground warfare includes a strong repository of Weapon-Target Allocation papers (which touch upon Weapon Control Warfare and Resource Allocation Warfare in the modern context), as well as papers that address ancient ground engagements and guerilla warfare. Where human lives are vulnerable, their protection is the most important element of these games, and the next priority is the protection of ground based assets.

\subsubsection{Sea (Naval) warfare}

Given the importance of navies for the projection of power globally, there is a surprising paucity of publicly available literature on naval warfare - with or without the application of game theory. There is often mention of naval warfare in papers dealing with target tracking, but a discussion on military naval strategy is limited to outdated literature or discussion of bare essentials \citep{harding2002seapower, speller2018understanding}. We will review the available papers in this regard, and highlight this an area where there is a sizeable gap in the literature.

\subsubsection{Aerial warfare}

It was not long after the Wright Brothers' invention of the aeroplane that aerial warfare became a critical factor in combat and military campaigns \citep{paris1992winged}. In a combat medium rarely impeded by obstacles or dimension, the nature of aerial combat is fast paced, intuitive and incredibly treacherous, with unpredictable `rules' for engagement \citep{colby1925laws, garner1924proposed}. In the present day, the factors to consider are vastly more complicated, and there is no shortage of resources available to military forces to conduct aerial combat- both human and machine \citep{gingras2000morality, dunn2013drones, stoyen2008method}. The literature shows that as a result of this abundance of arsenal, the intrinsic and potential value of both the targets and the resources used to engage is of particular importance in aerial warfare scenarios. Decisions about those values for both sides of the conflict need to be made when evaluating strategies for combat. As such, there are several papers which deal with the use of game theory in aerial warfare.

\subsubsection{Space warfare}

While the notion of warfare in space has existed for almost a century, neither physical execution nor established strategic theory of space warfare has been established \citep{klein2012space}. Nevertheless, this has not stopped military forces chasing the stars (literally and figuratively) \citep{wortzel2008chinese, johnson2016space}, and has inevitably lead to concepts from game theory being used in space warfare strategic thinking. This is currently mostly limited to satellite networks, where the key parameters of the game are optimised power use and signal strength across the network. The field is still quite young, and further military development in space seems to be inevitable, with which the corresponding literature dealing with applications of game theory in space warfare will also grow. 

\subsubsection{Cyber warfare}

 Cybersecurity is the protection of  IT systems and networks from being damaged/disrupted/subjected to information theft. Cyber warfare deals with the concept of Information and communication Systems being deliberately attacked to obtain a military advantage. While Cybersecurity has been an important field in computer science for many decades now, literature about cyber warfare as such is more scarce, and in any case heavily overlaps with applications of game theory in computer science in areas related to cybersecurity.  This review presents and analyses some papers which are specifically concerned with cyber warfare.

\subsubsection{Mixed/ other warfare}

There are several papers which address specific niches of defence applications of game theory, and yet cannot be classified as papers analysing a certain type of warfare. In some of these papers, the focus is more on the technology that is used: for example, target tracking.  In others, the nature of the hostile actors against whom the defence needs to be conducted changes: for example, national security operations which target domestic terrorism threats. There are several papers which deal with the use of game theory in such scenarios.

\textbf{Target tracking systems:} Target tracking in the military is the observation of a moving target and the surveillance of its position and manoeuvres \citep{blackman1986multiple,lipton1998moving, li2003survey}. Success in this domain relies on accuracy in the observed metrics and data, as well as efficient distribution and processing of all collected information \citep{kasthurirathna2014influence}. With the advent of intelligent targets, the military must also incorporate predictive methods to maintain ideal tracking performance. The literature reviewed  in tis regard covers topics from tracking strike missiles to theatre ballistic missiles, and tracking unknown intelligent agents to enemy aircraft. Key considerations in this area which shape the games played involve whether or not the target is `intelligent' / can take evasive action, weather or not target will have an optimal trajectory, and whether or not the target will have defenders \citep{bar1996estimation}. Target tracking applications of game theory mostly occur in aerial and naval warfare, including underwater surveillance.

 \textbf{National Security applications} : Game theory often finds application in national security and anti-terrorism related applications. This includes  predicting and preparing for terrorist attacks,  as well as resource allocation scenarios for protection of key personnel and landmarks /  other potential targets for terrorist activity.   While the value of potential targets and the likelihood of attacks are obviously key parameters governing the payoff functions of games in this niche,  the subsequent social, economic and political ramifications are equally instrumental  in modelling games  in this area \citep{gandhi2011dimensions, harris1992war}.  Few military conflicts have as much exposure as those on the home front \citep{maltby2007communicating}, and the fallout from terrorist attacks and their effect on public mood and confidence in the security apparatus are often taken into account in modelling payoff functions in this area.

\subsection{Game Theory}

Game theory, which is the study of strategic decision
making,  was first developed as a
branch of micro-economics~\citep{von1944game, Barough20121586, kasthurirathna2014topological, myerson1991game, von1928theory,  osborne2004introduction, von1944theory}. However, later it
has been adopted in diverse fields of study, such as
evolutionary biology, sociology, psychology, political
science, project management, financial management and computer science~\citep{rasmusen1994games, kasthurirathna2015emergence,kasthurirathna2013evolution,thedchanamoorthy2014influence, martin1978selective, piraveenan2019applications, peldschus2008experience, schotter2008economic, morrow1994game, hammerstein1994game, thakor1991game, myerson2013game}. Game theory
has gained such wide applicability due to the prevalence
of strategic decision making  scenarios across different disciplines. Game Theory provides insight into peculiar behavioural interactions like the cooperative interactions within groups of animals \citep{bshary2015cooperation, hadjichrysanthou2012should}, the bargaining and exchange in a marriage \citep{lundberg1994noncooperative, manser1980marriage,grossbard1995marriage} or the incentivisation of Scottish salmon farmers \citep{murray2014game}.  A typical game defined in game theory has two or more players, a set of strategies available to these players, and a corresponding set of payoff  values (sometimes called utility values) for each player (which are, in the case of two-player games, often presented as a payoff-matrix)\citep{rasmusen1994games, gibbons1992game, shubik1970game}.


\subsubsection{Pure vs Mixed Strategies}
A pure strategy in a game provides a complete definition of how a player will play a game. A player's strategy set is the set of pure strategies available to that player\citep{polak2007econ}.

A mixed strategy is a combination of  pure strategies where a particular probability $p$  (where $ 0 \le p \le 1$   ) is associated with each of these pure strategies. Since probabilities are continuous, there are infinitely many mixed strategies available to a player. A totally mixed strategy is a mixed strategy in which the player assigns a strictly positive probability to every pure strategy. Therefore, any  pure strategy is actually a degenerate case of a mixed strategy, in which that particular strategy is selected with probability $1$ and every other strategy is selected with probability $0$.  A totally mixed strategy is a mixed strategy in which   every pure strategy has a strictly positive (non-zero) value.

\subsubsection{Nash Equilibrium}
Nash equilibrium is one of the core concepts in game theory. It is a state (a set of strategies) in a strategic game from which no player has an incentive to deviate unilaterally, in terms of payoffs. Both pure strategy and mixed strategy Nash equilibria can be defined. A strategic game can have more than one Nash equilibrium~\citep{nash1950equilibrium, kasthurirathna2014optimisation}.  It is proven that every game with a finite number of players in which each player can choose from finitely many pure strategies has at least one Nash equilibrium~\citep{nash1950equilibrium}.

The formal definition of Nash equilibrium is as follows. Let $(S, f)$ be a game with $n$ players, where $S_i$ is the strategy set of a given player \textit{i}. Thus, the strategy profile $S$ consisting of the strategy sets of all players would be, $S = S_1$ x $ S_2 $ x $ S_3....$ x $ S_n$.  Let $f(x) = (f_1(x),....., f_n(x))$  be the payoff function for  strategy set $x\in S$. Suppose $x_i$ is the strategy of player \textit{i} and $x_{-i}$ is the strategy set of all players except player \textit{i}. Thus, when each player $i\in 1,.....,n$ chooses strategy $x_i$ that would result in the strategy set $x = (x_1,....,x_n)$, giving a payoff of $f_i(x)$ to that particular player, which depends on both the strategy chosen by that player ($x_i$) and the strategies chosen by other players ($x_{-i})$. A strategy set $x^*\in S$ is in Nash equilibrium if no unilateral deviation in strategy by any single player would return a higher utility for that particular player~\citep{kasthurirathna2015evolutionary}. Formally put,  $x^*$ is in Nash equilibrium if and only if:

\begin{equation}
\forall i,x_i\in S_i : f_i(x_i^*,x_{-i}^*)\geq f_i(x_i,x_{-i}^*)
\end{equation}\label{nash}

\subsubsection{Normal-form games and extensive-form games}

In a normal-form game, only a single round of decision making takes place, where all players make decisions simultaneously. An extensive-form game is, on the other hand,  an iterative game where there are several rounds of decision making~\citep{binmore2007playing, hart1992games}. On each round, players could make decisions simultaneously or in some pre-defined order.  An extensive-form game is often represented by a game tree, where each node (except terminal nodes) is a decision point, and each link corresponds to a decision or a set of decisions  that could be made by the relevant player / players at that point. The terminal nodes represent an end to the extensive-form game, with corresponding payoffs for each player involved.

\subsubsection{Non-cooperative games and cooperative games}

Typically, games are taken to be played for the self-interest of the players, and even when the players cooperate, that is because cooperation is the best strategy under the circumstances to maximise the individual payoffs of the players. In such games, the cooperative behaviour, if emerges, is driven by selfish goals and is transient. These games can be termed 'non-cooperative games'. These are sometimes referred to, rather inaccurately, as `competitive games'. Non-cooperative game theory is the branch of game theory that analyses such games. On the other hand, in a cooperative game, sometimes also called a coalitional game,  players form coalitions, or groups, often due to external enforcement of cooperative behaviour, and competition is between these coalitions~\citep{myerson2013game, nash1950equilibrium, nash1951non, ritzberger2002foundations, kasthurirathna2014optimisation}.  Cooperative games are analysed using cooperative game theory, which predicts which coalitions will form and the payoffs of these coalitions.  Cooperative game theory focusses on surplus or profit sharing among the coalition ~\citep{peleg2007introduction}, where the coalition is guaranteed a certain amount of payoff by virtue of the coalition being formed.  Often, the outcome of a cooperative game played in a system is equivalent to the result of a constrained optimisation process~\citep{bell2017network}, and as such, some  papers that we review use a linear programming framework to solve the cooperative games they model.

\subsubsection{Zero-sum games}
Zero-sum games are a class of competitive games where  the total of the payoffs of all players is zero. In two player games, this implies that one player's loss in payoff is equal to another player's gain in payoff. A two player zero-sum game can therefore be represented by a payoff matrix which shows only the payoffs of one player.  Zero-sum games can be solved with the mini-max theorem~\citep{Nash1953}, which states that in a zero-sum game there is a set of strategies which minimises the maximum losses (or maximises the minimum payoff)  of each player. This solution is sometimes referred to as a `pure saddle point'.   It can be argued that the stock market is a zero-sum game. In contrast, most valid economic transactions are non zero-sum since each party considers that, what it receives is more valuable (to itself) than what it parts with. 

\subsubsection{Perfect vs Imperfect Information Games}
In a  perfect information game, each player is aware of the full history of the previous actions of all other players, as well as the initial state of the game. In imperfect information games, some or all players do not have access to the entirety of information about other players' previous actions. 

\subsubsection{Simultaneous games and sequential games}

A simultaneous game is either a normal-form game or an extensive-form game where on each iteration, all players make decisions simultaneously. Therefore, each player is forced to make the decision without knowing  about the decisions made by other players (on that iteration). On the contrary, a sequential game is a type of extensive-form game where players make their decisions (or choose their strategies) in some predefined order~\citep{binmore2007playing, hart1992games}. For example, a negotiation process can be modelled as a sequential game if one party always has the privilege of making the first offer, and the other parties make their offers or counteroffers after that. In a sequential game, at least some players can observe at least some of the actions of other players before making their own decisions (otherwise, the game becomes a simultaneous game, even if the moves of players do not happen simultaneously in time). However, it is not a must that every move of every previous player should be observable to a given player. If a player can observe every move of every previous player, such a sequential game is known to have `perfect information'. Otherwise the game is known to have `imperfect information'.  Sequential games are often used by papers that we are reviewing here, to model bargaining or negotiation processes.

\subsubsection{Differential Games}
Differential Games are often extensive form games, but instead of having discrete decision points, they are modeled over a continuous time frame \citep{kamien2012dynamic}. In such games, the each state variable evolves continuously over time according to a differential equation. Such games are ideal for modelling rapidly evolving defense scenarios where each player engages in selfish optimisation of some parameter. For example, in missile tracking problems, the pursuer and the target both try to control the distance between them, whereas the pursuer constantly tries to minimise this distance and the target constantly tries to increase it. In such a scenario, iterative rounds of decision making are much too discrete to model the continuous movements and computations of each player.  Differential games are ideal to model such scenarios.

\subsubsection{Stackelberg games}

A Stackelberg game is a particular type of two player sequential game commonly used in economics ~\citep{von2010market}. In a Stackelberg game, there is a leader and a follower, which are typically companies operating in a market. The leader firm has some sort of market advantage that enables it to move first and make the first decision, and the follower's optimal decision depends on the leader's decision. If a follower chose a non-optimal action given the action of the leader, it will adversely affect not only the payoff of the follower, but the payoff of the leader as well. Therefore, the optimal decision of the leader is made on the assumption that the follower is able to see the leader's action and will act to maximise its own payoff given the leader's action. Several papers we review here have used Stackelberg games to model project management scenarios. Stackleberg games might be relevant in defence applications where there are leaders and followers, such as naval, airplane fighter or tank formations.

\subsubsection{Common interest games}

Common interest games are another class of non-cooperative games  in which there is an action profile that all players strictly prefer over all other profiles~\citep{calcagno2014asynchronicity}.  In other words, in common interest games, the interests of players are perfectly aligned. It can be argued that common interest games are the antithesis of zero-sum games, in which the interests of the players are perfectly opposed so that any increase in fortune for one player must necessarily result in the collective decrease in fortune for others.  Common interest games were first studied in the context of cold war politics, to understand and prescribe strategies for handling international relations~\citep{schelling1958strategy, schelling1980strategy, lewis2008convention}.

It makes sense to classify non-cooperative games into common interest games and non-common interest games, just as much as it makes sense to classify them into zero-sum games and non-zero sum games, since these two concepts (zero-sum games and common interest games) represent extreme cases of non-cooperative games. However, the papers that we review do not use common interest games to model project management scenarios, and indeed it would be rare to find scenarios in project management where the interests of players are perfectly aligned. Therefore,  we do not use the common interest games-based classification in our classification process, as it would add another dimension  and increase the complexity of classification needlessly.

\subsubsection{Nash bargaining}

In a Nash bargaining game~\citep{nash1950bargaining, myerson1999nash}, sometimes referred to as a bargaining problem or bargaining game,  two players could choose from an identical set of alternatives, however each alternative has varying payoffs for the players. Typically, some alternatives have better payoff for one player, while other alternatives have better payoff for the other player. If both players choose the same alternative, then each get the payoff corresponding to that alternative. If they choose differing alternatives, then there is no agreement, and they each get a fixed payoff which corresponds to the cost of non-agreement, and therefore typically negative.  Thus, there is incentive to choose an alternative which may not necessarily be the best  for a player. If there is perfect information, that is, the full set of alternatives and payoffs is known to both players, then there is an equilibrium solution to the Nash bargaining game~\citep{rubinstein1982perfect}.

\subsubsection{Subgames}

A subgame is a subset of a sequential game such that at its (the subgame's) beginning, every player knows all the actions of the players that preceded it~\citep{osborne2004introduction}.  That is, a subgame is a section of the game tree of a sequential game where the first decision node that belongs to this section has perfect information.

\subsubsection{Subgame perfect Nash equilibrium}

In a sequential game, a subgame perfect Nash equilibrium is a set of strategies representing each player such that they constitute a Nash equilibrium for every subgame of that sequential game ~\citep{osborne2004introduction}. Thus, when a game tree of the sequential game is considered, if a set of strategies could be identified so that they represent a Nash equilibrium for every branch of game tree originating from a node which represents a point where every player knows all preceding actions of all players, such a set of strategies represent a subgame perfect Nash equilibrium for that sequential game. For example, when two players are bargaining, they are in a subgame perfect Nash equilibrium if they are presently employing a set of strategies, which will represent a Nash equilibrium between them at any future stage of the bargaining process  given that they are aware of the full history of the bargaining process up to that point.

\subsubsection{Behavioural Game Theory}

Behavioural Game Theory combines classical game theory with experimental economics and experimental psychiology, and in doing so, relaxes many simplifying assumptions made in classical game theory which are unrealistic. It deviations from several simplifying assumptions made in classical game theory such as perfect rationality\citep{camerer2011behavioral,kasthurirathna2015emergence},  the independence axiom, and the non-consideration of altruism or fairness as motivators of human decision making\citep{aumann1992handbook, aumann2008game}. We will show in this review that the approaches related to behavioral game theory are crucial in modelling military scenarios, such as in signaling games.

\subsubsection{Evolutionary game theory}

Evolutionary game theory is an outcome of the
adoptation of game theory into the field of evolutionary
biology ~\citep{smith1993evolution}.   Some of the critical questions asked in
evolutionary game theory include: which populations/strategies are stable? which strategies can `invade' (become popular) in populations where other strategies are prevalent? How do players
 respond to other players receiving or perceived to be receiving better payoffs in an iterated game setting? etc. Evolutionary games are often modelled as iterative
games where a population of players play the same game iteratively in a well-mixed or a spatially distributed
environment ~\citep{le2007evolutionary, kasthurirathna2015evolutionary}.

A strategy can be identified as an evolutionary stable strategy (ESS) if, when prevalent,  it has the potential to prevent any mutant strategy from percolating its environment~\citep{kasthurirathna2015evolutionary}. Alternatively, an ESS is the strategy which, if adopted by a population in a given environment, cannot be invaded by any alternative strategy. Hence, there is no benefit for a player to switch from an ESS to another strategy. Therefore, essentially, an ESS ensures an extended Nash equilibrium. For a strategy $S_1$ to be ESS against another `invading' strategy $S_2$, one of the two conditions  mentioned below needs to be met,  in terms of expected payoff $E$.

\begin{enumerate}
	\item $E(S_1,S_1) > E(S_2,S_1)$ : By unilaterally changing strategy to $S_2$,  the player will lose out against another player who sticks with the ESS $S_1$.
		\item $E(S_1,S_1) = E(S_2,S_1)$ \& $E(S_1,S_2) > E(S_2,S_2)$ :  a player, by converting to  $S_2$, neither gains nor loses against another player who sticks with the ESS $S_1$, but playing against a player who has already `converted' to $S_2$, a player is better off playing the ESS $S_1$.
\end{enumerate}

If either of these conditions are met, the new strategy $S_2$ is incapable of invading the existing strategy $S_1$, and thus, $S_1$ is an ESS against $S_2$. Evolutionary games are typically modelled as iterative games, whereby  players in a population play the same game iteratively.

\subsubsection{Signaling Games}
A signaling game \citep{cho1987signaling} is an incomplete information game where one player has perfect information and another does not. The  player with perfect information (the Sender \(S\)) relays messages to the other player (the Receiver \(R\)) through signals, and the other player will act on those signals after inferring the information hidden in the messages. The Sender \(S\) has several potential \textit{types}, of which the exact type \(t\) in the game is unknown to the Receiver \(R\). \(t\) determines the payoff for \(S\). \(R\) has only one type, and their payoff is known to both players.

The game is divided into the sending stage and the acting stage. \(S\) will send one of \(M = \{m_{1}, m_{2}, m_{3}, ..., m_{j}\}\) messages. \(R\) will receive that message and respond with an action from a set \(A = \{a_{1}, a_{2}, a_{3}, ..., a_{k},\}\). The payoff that each player receives is determined by the combination of the Sender's type and message, as well as the action that the Receiver responds with. An example of the signaling game is 
the Beer-Quiche Game \citep{cho1987signaling}, in which Player B, the receiver, makes a choice of whether or not to duel Player A. Player A is either surly or wimpy, and Player B would only like to duel the latter. Player A chooses to have either a Beer or a Quiche for breakfast. While they prefer a quiche, a quiche signals information from a stereotype that quiche eaters are wimpy. Player B must analyse how each decision, duel or not duel, will give them payoff depending on which breakfast Player A chooses.

With this background, we now review the available literature which deals with the applications of game theory in defence science and technology.


\section{Use of game theory in defence science and technology applications} \label{analysis}

As mentioned earlier, the primary parameters that influence the payoff matrix in games modelling defence scenarios are the value of targets, the value of resources and the priority of objectives. Other than this, the games used in defence applications can vary greatly, as we will see below. For this reason, this section is structured based on the domain (type of warfare) each paper covers Where a paper covers more than one domain, it is included in the most relevant subsection / domain. We however analyse in detail the type of games used, the way the payoff functions were structured, the available strategies and equilibria etc for each paper.

\subsection{Papers dealing with Land Warfare}

In Land warfare related applications of game theory, most studies focus on defensive warfare, whereby  the military makes decisions on how to best allocate their ground defences to multiple threats.  Some studies also focus on historical land-based conflicts and provide game theoretical analysis in hindsight, revealing how some decisions made from intuition in historical conflicts  had a rational and mathematical justification. Land warfare can result in very heavy casualties, so understanding how to best minimise human losses is a key component (though not the only objective) of  land warfare. Quite often, prioritising military resources is also  fundamental to  success and often features prominently in strategic decisions.  Furthermore, often in scenarios involving ground warfare, it is important to assess the knowledge about  opponents , their possible tactics, or terrain: it may become necessary  to combat airborne forces being inserted at certain places, or it may be needed to traverse uncertain territory. In each of these situations, understanding where a force has  imperfect information will help that force to make  rational decisions.

Bier et al. \citep{bier2008} design a game to best assign defensive resources to a set of locations / resources that need to be protected. The attacker must then decide how they choose to split their force to attack the different targets. The game is modelled as a two player game of normal form. Payoff in this game is absolute, and an attack on a location \(i\) is either a success or a failure, where the attacker gains \(a_{i}\) and the defender loses \(d_{i}\). Since orders for an attack are confirmed ahead of an attack, attackers must use a set of pure  strategies. The game can be played both simultaneously or sequentially. That is, the game can be played depending on whether or not the attacker knows how the defender has assigned their resources before making their decision. This leads to the ideal strategy being to leave some targets undefended, centralising defenses in key areas by  leaving some areas vulnerable.

There are several studies which analyse historical conflicts, which occurred predominantly on land, using a game theoretic prism. For example,  Cotton and Liu's \citep{cotton2011100} describe two ancient Chinese military legends and model them as signaling games. In both games, legendary military leaders are faced with formidable opposing armies with much greater numbers and strength than their own, but instead of retreating, they prepare to engage, acting as if they are setting up for an ambush. Their opponents with imperfect information are left only with the messages they can infer from their opponents' actions; spooked by the perceived confidence and the reputation that these generals carried, the opposing armies, though in actuality are of superior strength, choose not to engage. Through a brave and ingenious bluff, both generals achieve an equilibrium solution in their favour by standing their ground. They do this by creating deception without direct communication, which follows the template of the aforementioned Beer-Quiche signaling game.

The first game described by Cotton and Liu is the "100 Horsemen" game. They describe a piece of history where a hundred Han horsemen travelling alone encounter a large Xiongnu force numbering in the thousands. Their available strategies are to retreat or engage. If they retreat, and the enemy engage, they will very likely be run down and defeated; and if they engage and the enemy also engage, they will be eliminated in battle. The best outcome for them is to somehow force an enemy retreat. The enemy is uncertain if the horsemen are travelling with a greater army. They see the horsemen move to engage, and decide not to take the risk,  and retreat. The situation is translated into a two player game, with two strategies. It is represented in Figure \ref{fig:100horse} below:

\begin{figure*}[htbp]
\centering
\includegraphics[width=0.98\textwidth]{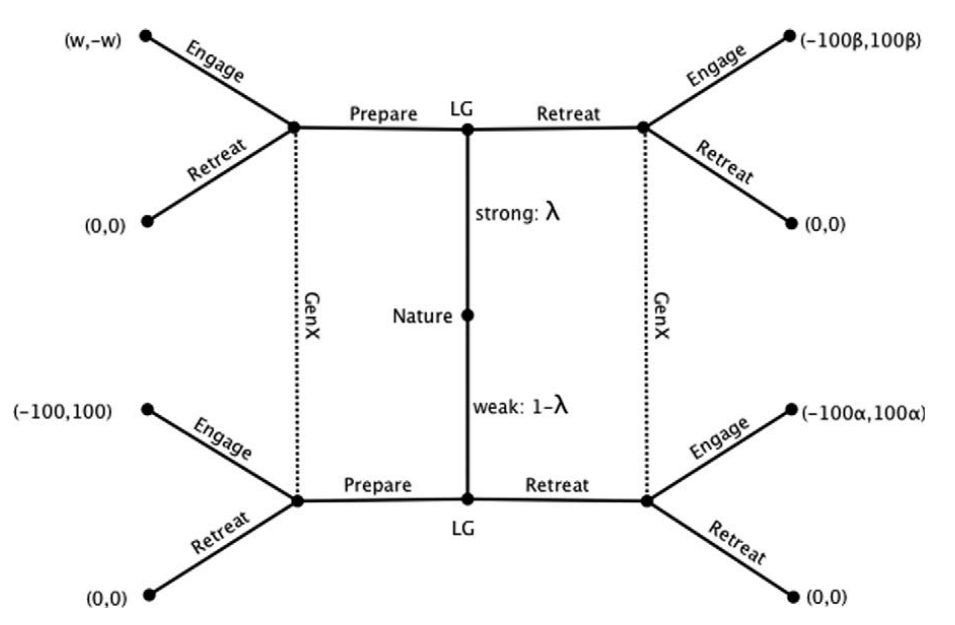}
\caption{The 100 Horsemen signaling game \citep{cotton2011100}}
\label{fig:100horse}
\end{figure*}

In Figure \ref{fig:100horse} :

LG represents the decision point for the General Li Guang, of the Han forces.

GenX represents the decision point for the opposing Xiongnu force.

Payoffs are listed as (LG, GenX)

\(\lambda \in (0,1)\) represents the General's ability, as either strong or weak

\(\alpha\) and \(\beta\) represents the proportion of Han horsemen killed in a retreat 

\(w\) is a positive parameter\newline

The second game is very similar to the first.  In this game, a small city is guarded by the formidable General Zhuge Liang. He learns that a great  hostile army is approaching the city. He is faced with two options. He could run, after which he would secede the city and likely be chased down by the approaching army, or he could stay and defend the city. If he chose the latter, and the army were to engage, he would likely lose his life, his army and the city. Faced with this dilemma, he orders his men to hide out of sight, so that the city appears empty from the outside. He climbs to the top of the foremost tower of the city and plays music. The opposing general, aware of General Liang's experience and prowess, suspects that the General has taken this unassuming position in the tower in the empty city to ambush his army, and they move away from the city to avoid being ambushed. General Liang sent effectively two signals here. The first was his reputation, a signal encompassing his strategic and military strength. The second was his choice to stay and defend the city. With these two pieces of information, and nothing else about the whereabouts or magnitude of General Liang's army, the opposing army chooses the safe option of zero loss and leaves. This piece of history is modelled as another two player signaling game, shown in Figure \ref{emptycity} below:

\begin{figure*}[htbp]
\centering
\includegraphics[width=0.70\textwidth]{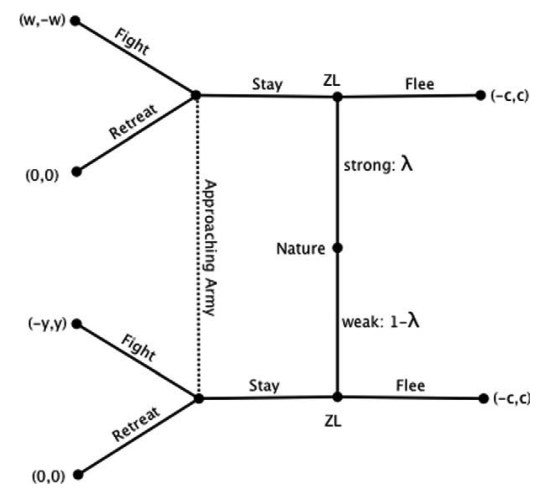}
\caption{The Empty City signaling game \citep{cotton2011100}}
\label{emptycity}
\end{figure*}

In Figure \ref{emptycity} :

ZL represents the decision point for the General Zhuge Liang

Payoffs are listed as (ZL, Opposing Army)

\(\lambda \in (0,1)\) represents the General's ability,  as either strong or weak

\(c\) represents the value of the city

\(w\) represents the gains if ZL's army matched the opposing army

\(y\) represents the losses if ZL's army is weaker than the opposing army, and \(y > c\) since it encompasses losing the city\newline

Both pieces of history  represent distinguished military decision making in the face of near certain defeat, and  are in fact examples of Generals with  a strong understanding of the nuances of signals and rational decision making in strategic interactions forcing a favourable outcome to themselves.

The next paper we review is Gries et al \citep{gries2016economic}  which is a comprehensive investigation into the utility of game theory principles in guerilla/destabilisation warfare.  The significant  factors they model  are: destabilisation insurgents often attack randomly, creating a continuous threat that must have a continuous mitigation and detection strategy; duration of a war is important to consider, and will change the value that is assigned to targets and assets; time preferences play a critical role in setting priorities, as always judgements of value determine strategic decisions which in turn  determine success or failure. The game  model they propose involves both a sequential  non cooperative game and a simultaneous non cooperative game,  in each of which the two  players are the guerilla force and the government. For these conflicts, the economic and social impacts  are much more significant than military losses and gains, and therefore play a much more significant role in calculating the value of outcomes.

\begin{figure*}[htbp]
\centering
\includegraphics[width=0.70\textwidth]{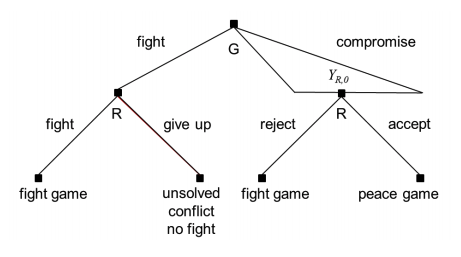}
\caption{Destabilisation Warfare game \citep{gries2016economic}}
\label{guerilla}
\end{figure*}

The game specifically looks at moments when each side looks to try and find peace or conflict with the other. At each of these moments, the government force must consider the financial cost of each option, while the rebels must on the order of  priority of the engagements,  and what portion of their fighting force they will make available for each engagement. Figure \ref{guerilla}  demonstrates an example of the decision tree to emerge from these moments in destabilisation warfare, where G represents the Government decisions and R represents the Rebel decisions. 

Krisnamurthy et al. \citep{krishnamurthy2007game} investigate a game-theoretic control of dynamic behaviour of an unattended ground sensor network (UGSN) in order to acquire information about intruders. Each sensor in this network is capable of receiving measurements, with specified accuracy, of the range and bearing of nearby targets which they then transmit to a local hub for data fusion. In this framework, while more sensor measurements and larger volumes of transmission of measurements may lead to better target awareness, this also results in the undesirable effect---greater consumption of limited battery power. Hence, the goal to which game-theory is applied is to optimally trade-off target awareness, data transmission and energy consumption using a two-time scale, hierarchical approach. 

The authors demonstrate that the sensor activation and transmission scheduling problem can be decomposed into two coupled decentralized algorithms. In particular, the sensors are viewed as players in a non-cooperative game and an adaptive learning strategy is proposed to activate the sensors according to their proximity to targets of interest. This turns out to be a correlated equilibrium solution of this non-cooperative game. Next, the transmission scheduling problem, in which each sensor has to decide at each time instant, whether to transmit data and waste battery power, or to wait and increase delay, is formulated as a Markov Decision process with a penalty terminal cost. The main result of this formulation is to show that the optimal transmission policy has a threshold structure which is then proved using the concept of supermodularity.\\

\subsection{Papers dealing with Naval  Warfare}

Naval warfare predates aerial warfare and has been prevalent for a longer duration in human history. In the 18th and 19th centuries, the powerful nations of the time built warships with cannons positioned along their sides. It meant that ships could attack typically only to their sides. When sailing as an armada, the standard approach was to form a `line of battle' i.e. a column of allied naval ships sailing in a direction such that their sides would face the enemy, also positioned in a line. The two parallel opposing fleets could blast one another with large number of cannons. The `line of battle' strategy is considered to be a Nash-equilibrium because neither fleet would gain from performing raking (a tactic of the era, whereby an attacking ship would attempt to sail across its adversary's stern, concentrating cannon fire there while the enemy could only respond minimally due to having less cannon defences in the stern. The attacking ship would damage both the stern and some of the broadsides of its adversary). According to \citep{levine1995pacific} 
raking was not preferred in a fleet as this would mean having to first sail ahead of its enemy and then turn towards it---a challenging task when the ships' speeds were roughly equal and maneuvering was difficult. As neither fleet would gain from turning towards the enemy and neither would get ahead, Levine concludes that this strategy---forming a line of battle and sailing parallel to the other fleet---was each fleet's best response, and thus represented a Nash equilibrium. 

Levine goes on to mention battles in which English fleets deviated from the above strategy and sailed orthogonally towards a French and Franco-Spanish fleet. In the first battle Levine mentions, it was likely unplanned. In the second---the 1805 Battle of Trafalgar---it was by careful design: the English fleet divided itself into two columns, each of which sailed orthogonally to the Franco-Spanish line, raking fire for about 45 minutes before crashing through it and beginning a general melee. The English would go on to isolate the middle of the Franco-Spanish fleet to score a decisive victory. Levine considers both battles to be counterexamples to his thesis. However, in the Battle of Trafalgar, its is possible that the English strategy was a best response to the likely Franco-Spanish strategy of forming an orthodox line of battle. The English admiral, Lord Nelson, desired to keep the Franco-Spanish fleet from escaping---which they could if both fleets formed parallel lines of battle---thus reducing the reward he would get for forming his own fleet into a line of battle. Moreover, he may have estimated that the poor gunnery of the French and Spanish ships would lessen the effect of the raking fire, thus reducing the negative reward he would get for directly charging the Franco-Spanish fleet. In his eyes, this may have made the unorthodox option a better response to the likely Franco-Spanish strategy than the orthodox line of battle. While Levine did not explicitly attribute these strategies in naval battles of the era to game theory, it is quite possible that the adopted strategies had a game-theoretic basis.

Maskery et al. \citep{maskery2007network} study the problem of deploying counter-measures against anti-ship missiles using a network enabled operations (NEOPS) framework, where multiple ships communicate and coordinate to defend against a missile threat. Here, the missile threats are modelled as a discrete Markov process and they appear at random positions within a fixed physical space and move towards the ships obeying some known target dynamics and guidance laws. The ships which are equipped with counter-measures (CM) such as decoys and electromagnetic jamming signals, are modelled as the players of a transient stochastic game, where the actions of the individual players include the use of CM to maximize their own safety while cooperating with other players which are essentially aiming to achieve the same objective. The optimal strategy of this game-theoretic problem is a correlated equilibrium strategy, and is shown to be achieved via an optimization problem with bilinear constraints. This is in contrast to the Nash equilibrium solution proposed in \citep{maskery2007decentralized} to a related problem but one without player coordination. A noteworthy contribution of this paper is that it also quantifies the amount of communication necessary in order to implement the NEOPS equilibrium strategy. This paper highlights the utility of game-theoretic methods in analysing optimal strategies in network enabled systems which are critical in modern warfare.\\

In \citep{maskery2007decentralized} Maskery et al. consider the problem of network centric force protection of a task group against anti-ship missiles. The decision makers in this model are the ships equipped with hard-kill/soft-kill weapons (counter-measures) and these ships are also considered the players in the formulation of this problem as a game-theoretic setting. The platforms must make critical decisions independently on the optimal deployment of the counter-measures while they simultaneously work towards a common goal of protecting the members of the task group. Essentially this is a decentralised missile deflection problem in a naval setting which is formulated as a transient stochastic game for which the ships may compute a joint counter-measure policy that is in Nash equilibrium. Here the ships play a game with each other instead of with a missile. This approach naturally lends itself to decentralised solutions which may be implemented when full communication is not feasible. Moreover, this formulation leads to an interpretation of the problem as a stochastic shortest past game for which Nash Equilibrium solutions are known to exist.\\

Bachmann et al. \citep{bachmann2011game} analyse the interaction between a radar and jammer using a noncooperative two-player, zero-sum game. In their approach, the radar and jammer are considered ``players'' with opposing goals: the radar tries to maximize the probability of detection of the target while the jammer attempts to minimize its detection by the radar by jamming it. Bachmann et al. \citep{bachmann2011game} assume a Swerling Type II target in the presence of Rayleigh distributed clutter, for which certain utility functions are described for cell-averaging (CA) and order-statistic (OS) CFAR processors in different cases of jamming. This game-theoretic formulation is solved by optimizing these utility functions subject to constraints in the control variables (strategies), which for the jammer are jammer power and the spatial extent of jamming while for the radar the available strategies include the threshold parameter and reference window size. The resulting matrix-form games are solved for optimal strategies of both radar and jammer from which they identify conditions under which the radar and jammer are effective in achieving their individual goals.\\

\subsection{Papers dealing with Aerial Warfare}

Aerial combat is often a normal-form  game where decisions about utilised resources are made before the engagement, based on assumptions and knowledge about the strength of different elements of the arsenal. For example, Suppression of Enemy Air Defense vehicles (SEADs) are effective against ground-to-air defenses and Surface to Air Missiles (SAMs), but will not be useful against fighter aircraft. Therefore, when military personnel decide which resources to use in an engagement, they need to weigh how valuable each of their resources is, as well as how important the objective is to both sides of the conflict. If the attacking force values a target much more than it is actually worth, then their increased resource expenditure may be detrimental to their military campaign as a whole. With humans operating the aerial weaponry usually, their respective abilities and skill sets need to be considered, and the  likelihood of them executing their mission.

There is limited literature on aerial combat modelled with  game theory.  Hamilton \citep{hamilton2004simple} writes a comprehensive guide to applications of game theory on multiple Aerial warfare situations. Hamilton suggests using game theory to devise strategies not only based on one's own military options, but also expectations around enemy actions as well. Game theory accounts for different interactions with the enemy, rather than  simply considering which side had superior maximum-effort power. Nowadays, many military forces are able to adapt to instantly changing situations and adjust their actions based on those new circumstances. As such, Hamilton suggests first determining all of the tactical options available to each side. As stated earlier, one of the most fundamental elements of using game theory for the military is understanding exactly how much value each asset holds - and detailing the inventory and strategic possibilities of both sides will best clarify all strategic options. For each option, Hamilton suggests assigning a numerical value - a Measure of Effectiveness (MoE). Decisions about MoEs are important because being accurate with MoEs will underpin the choices that are made strategically. Incorrect MoEs can lead to incorrect strategic decisions, and perhaps also resulting in poor understanding  of why the decision was wrong. An example of this  (although not in the aerial warfare context) was the Vietnam War, where the early US strategy was maximising the neutralisation of Viet Cong soldiers. Since the Northern Vietnamese leadership did not place great emphasis on their infantry, the US strategy ultimately led to a loss in the war.  Next, Hamilton suggests calculating  the combined value for all possible interactions between strategies of both sides of the conflict. This will generate a matrix of payoffs, from which it is possible to derive the optimum or dominant strategy for each player, and then  an equilibrium solution. Thus, ahead of any engagement a military leader may partake in, they have a well-formed idea of the expected result of the game. A caveat that Hamilton adds to these guidelines is to consider the length of a military campaign as a whole. The values that can be assigned to a resource for one battle or strike attack may be small if they are low cost or numerous asset, but depending on how many of these skirmishes happen over the course of a campaign, those resources may be come pivotal over the course  of a war. 

To illustrate these points, Hamilton applies them to a standard aerial warfare game of SEADs and time critical targets. In this combat, the 'Blue' side is trying to eliminate some ground based targets. To do this, they use SEADs. In response, the 'Red' side will fire SAMs, which SEADs struggle to avoid. However, expecting this response, the Blue side also has Strike aircraft which can defend the SEADs and counteract the SAMs, but are unable to attack the targets. The questions for the Blue team are: what is the value of the target and what ratio of SEADs and Strike aircraft should be deployed for the targets? Likewise, for the Red team: how valuable is the target and how many, if any SAMs should be fired? Hamilton contends that the optimal Red strategy is to fire only for a fraction of engagement which is equal to:
\begin{equation}
    \frac{\text{Value of Target}}{\text{Value of Target} + \text{Value of SEAD } \times Pk_{s} + \text{Value of SAM } \times Pk_{A}}
\end{equation}

and the optimal Blue Strategy is to assign a fraction of the planes as SEADs which is equal to:

\begin{equation}
    \frac{\text{Value of SAM } \times Pk_{A}}{\text{Value of SAM } \times Pk_{A} + \text{Value of SEAD } \times Pk_{s} + \text{Value of Target}}
\end{equation}
where 

\(Pk_{s}\) is the probability of the SAMS destroying the SEADs

\(Pk_{A}\) is the probability of the Strike aircraft destroying the SAMs\newline

This formulation gives a concise prediction of the likely outcome of an engagement given every possible assignment of aircraft and missile launches. It must be noted that it is incredibly difficult in practice to accurately quantify the numerical value of different targets and resources.

Deligiannis et al. \citep{deligiannis2016power} consider a competitive power allocation problem in a MIMO radar network in the presence of multiple jammers. The main objective of the radar network is to minimize the total power emitted by the radar while achieving a specific detection criterion for each of the targets. In this problem, the radars are confronted by intelligent jammers that can observe the radar transmitted power and thereby decide its jamming power to maximize interference to the radar. Here they treat this power allocation problem as a non-cooperative game where the players are the central radar controller and the jammers, and solve this using convex optimization techniques. Moreover, they provide a proof for the existence and uniqueness of the Nash equilibrium stable point, where no player can further profit by changing its power allocation.\\

Garcia et al. \citep{garcia2019strategies} investigate the problem of defending a maritime coastline against two enemy aircraft whose main objective is to invade the territory controlled by the defending aircraft. The defender, on the other hand, attempts to prevent this by trying to intercept both enemy aircraft in succession as far as possible from the border. This is a typical pursuit-evasion scenario and is representative of many important problems in robotics, control and defence. In this paper, Garcia et al. formulate this problem as a zero-sum differential game, where the defender/pursuer tries to successively capture the two attackers/evaders as far as possible from the defended coastline while the attackers cooperate and minimize their combined distance from the border before they are captured. They then find the optimal strategies for the attackers and the defender in this one-defender two-attacker pursuit-evasion game by solving a set of nonlinear equations. The cooperative strategy discussed in this paper provides an important coordination approach for less capable (perhaps slower) agents when they are tasked to carry out a mission.\\

He et al \citep{he2019joint} consider the radar counter-measure problem in a multistatic radar network, where a game-theoretic formulation of joint power allocation and beamforming is studied in in the presence of a smart jammer. The goal of each radar in this network is to meet the expected detection performance of the target while minimizing its total transmit power and mitigate the potential interferences. On the other hand, the goal of the jammer is to adjust its own transmit power to interfere the radar so as to protect the target from detection. First, they study the power allocation game with strategy sets of each player (radar and jammer) consisting of their respective transmit powers. For this problem, they proceed to solve the corresponding optimization problems to work out the best response function for the radar and the jammer and show the existence and uniqueness of the Nash equilibria. Next, they consider the joint power allocation and beamformer design problem in the presence of jammers again as a non-cooperative game and propose a power allocation and beamforming algorithm which is shown to converge to its Nash equilibrium point.\\

McEneaney et al. \citep{mceneaney2004stochastic} investigate the command and control problem for unmanned air vehicles (UCAVs) against ground-based targets and defensive units such as surface-to-air missile (SAM) systems. The motivation for this work arises from the requirement for operations planning and real-time scheduling in an unmanned air operations scenario. The problem is modelled as a stochastic game between blue players (UCAVs) and red players that comprise the SAMs and ground based targets. The game objective may vary, an example of which for a blue player is to destroy a strategic target while minimizing damage to itself. The red players, on the other hand, attempt to inflict maximum damage on the UCAV while protecting themselves from attack by the UCAVs.

The control strategies for the UCAVs consist of a set of discrete variables that correspond to the specific target or SAM to attack while that for the SAMs are to switch their radar ``on'' or ``off''. Note that when the radar is ``on'', the probability of the SAM causing damage to the blue players increases as does the probability that the blue players inflict damage on the SAM. The solution to this stochastic game is obtained via dynamic programming and illustrated on some numerical examples. A main contribution of this work is the analysis of a risk-sensitive control based approach for stochastic games under imperfect knowledge. In particular, this approach not only handles noisy observations due to random noise, but also deals with cases that include an adversarial component in the observations.\\

Wei et al. \citep{wei2009hybrid} have developed a mission decision making system for multiple uninhabited combat aerial vehicles (UCAVs) working together. The UCAVs weapons are air-to-air missiles. In the paper a red-UCAV team consisting of an unmanned fighter-bomber flanked by two UCAVs attempts to strike a blue-team ground target. The blue-team has its own set of UCAVs that are directed to defeat the red-team. The success of a given missile against its chosen threat is determined by the distance between the attacker and threats, their relative speed, and relative angles. The blue team versus red-team scenario is represented as a simultaneous normal form game with the strategies for the team corresponding with allocations of blue team entities against red-team entities and vice versa. In the paper the payoff for red or blue team is based upon considering the effectiveness and ineffectiveness of a given allocation, which is turn is dependant upon the relative geometry between the opposing team allocation groupings. Dempster-Shafer (D-S) theory is applied where the D-S combinatorial formula is harnessed to formulate the payoff. These payoffs, calculated for each strategy, for each team is then placed into bi-matrices i.e., one for each team and solved using a linear programming optimisation approach. If an optimal Nash equilibrium point is not present, mixed strategy approach is applied and solved for. The authors have developed some mission scenarios with differing geometries and illustrated the use of their game-theoretical allocation strategy. They use annotated diagrams of entity geometry containing red and blue teams in proximity to one another to justify that the allocation strategy determined by their payoff formulation is satisfactory. \\

Ma et al. \citep{ma2019cooperative} have developed a game-theoretic approach to generate a cooperative occupancy decision-making method for multiple unmanned aerial vehicle (UAV) teams engaged against each other in a beyond-visual-range (BVR) air combat confrontation. BVR combat is targeted because of developments in missile technology enabling long-range engagements. In the paper, the team on each side first decides the occupancy positions (cubes in Cartesian space) of its UAV entities followed by selecting targets for each UAV team member to engage. The goal is for each side to obtain the greatest predominance while experiencing the smallest possible threat condition. A zero-sum simultaneous bi-matrix game is applied to express the problem. For a given occupancy of a UAV, height and distance predominance formulas that factor in range and weapon minimum/maximum performance criteria are used to generate payoff values for the utility functions. Since, the scale of the game leads to a explosion in size (and thus strategy) as the number of occupancy cubes and UAVs for each team is increased, the authors have chosen to augment the Double Oracle (DO) algorithm, that was designed to solve large scale zero sum game problems in earlier works by combining it with a neighbourhood search (DO-NS) algorithm. Through simulations, the authors illustrate that the results show the DO-NS algorithm outperforming the DO algorithm in terms of computational time and solution quality.\\

Evers et al. \citep{evers2013cooperative} have developed the design for a Cooperative Ballistic Missile Defence Game (CBMDG). This game has been designed to support strategic negotiations between a threatened nation and a possible coalition of nations that can be integrated together to form a layered BMD to protect the threatened nation. Through the game, the assignment of ballistic missile interceptors to the coalition nations is determined which minimizes the expected number of interceptors required to achieve a desired defence level in case of an attack from an enemy. The paper shows that the number of interceptor required by a coalition is smaller compared to utilising interceptors solely within the threatened nation to engage the BM threat, hence motivating it to seek out a coalition based layered defence. For the negotiation, the cost savings that threatened nation gains due to having to spend less on interceptors by leveraging cooperation via a coalition corresponds with the bargaining power that the nations in the coalition can demand as compensation. Using the interceptor savings game formulation devised in the paper, the savings introduced through each nation participating in the coalition can be determined and the fair compensation can be allocated to each aiding participating nation.

Garcia et al. \citep{garcia2017cooperative} consider an air combat scenario where a target aircraft that is engaged by an attacking missiles utilises a defending missiles as a countermeasure to defend itself as it attempts to escape the attacker by maximising my the distance between itself and the attacker when the defender reaches as close at it can to the attacking missile. The game is referred to as an  active target defense differential game (ATDDG). In the paper, the authors extend previous works performed on this three party problem to develop a closed-form analytical solution for the ATDDG where the Defender missile is able to defeat the attacker if it enters within a with a capture circle with specified radius rc > 0. Additionally, the closed-form optimal state feedback solution generated in the paper is supposed to work in spite of the attacker employing an unknown guidance law rather than assuming it is Propotional Navigation (PN) or pursuit (P). Finally, the authors provide the set of initial conditions for the target aircraft where its survival is guaranteed if the target-defender team plays optimally depsite the unknown guidance law employed by the attacking missile. \\

Han et al. \citep{han2016game}  present an Integrated Air and Missile Defence (IADS) problem where surface-to-air (SAM) batteries equipped with interceptor missiles (IM) engage the attacker missiles (AM) targeted at damaging cities being protected. The problem is cast as a simplified two-party zero sum (equally valued targets) game with perfect information and has three stages. By perfect information, it is implied that both players know what has happened in the previous stages of the game  The three stages correspond with defender setting up their allocation of SAMs to cities, followed by attacker allocating their missile salvo against cities and finally the defender in response allocating interceptor missiles to counter attacker missile salvo.  The simplifying assumptions made in this problem are that there is only one SAM allocated near a city and only one installed per site. Additionally, no more than one interceptor is launched against each attacking missile. Additionally, only one IM can be allocated to one DM, each each SAM has the same number and type of IMs, AMs are identical and are fired in a single salvo. It is attempted to be solve the tri-level game using an extensive form game tree,  $\alpha-\beta$ pruning and using the Double Oracle (DO) algorithm for a 6 city network that needs to be protected. Only DO is a heuristic based approach and is not guaranteed to find the SPNE. The efficiency with which sub-game Nash equilibrium is reached by each choice of algorithm is studied. For the game-tree approach, the conclusion made is that  size of the strategy space is determined to increase to an intractable size because of the combinatorial  nature of the problem. When applying  $\alpha-\beta$ pruning, the author determines that the when scaling to change number to increase SAM batteries, AMs and IMs, it does not scale well in terms of computation time compared to DO. However DO does fail to find the SPNE in a small number of instances. The author preferences DO despite its lack of guarantee to reach the SPNE since it is shown to not violate monotonicity (increase in payoff) and the solution quality trends (non exponential increase in computational time) even when increasing the size of the problem from 6 cities to 55 cities.

The work of Ba{\c{s}}p{\i}nar, Bar{\i}{\c{s}} et al. \citep{bacspinar2020assessment} focuses on modeling of air-to-air combat between two unmanned aerial vehicles (UAVs) using an optimization-based control and game-theoretic approach. In this work, the ability to provide a controller despite the presence of complex non-linear dynamics for a UAV is achieved through avoiding integration of the differential equations involved. It is possible to do this because the differential flatness theory can be successfully harnessed to describe the movements of the UAV in the horizontal and vertical planes. In particular, vehicle motion is expressed in terms of specific variables and their derivatives through parameterized curves. Then, any trajectory planning, for moving from one waypoint to another can be solved by determining the smooth curves satisfying defined conditions in flat output space. Following determination, all variables involved to describe the smooth curve can be reverted to the original state/input space. The impact is a speed-up in the solving of any trajectory optimization through reducing the number of variables required. Game theory is then harnessed where the aerial combat between the two UAVs is modelled as a zero-sum game using a minimax approach. That is the each party tries to maximise its payoff when the opponent plays its best strategy. Here, The objective is for each UAV is to get directly behind the other party and within a range threshold in order to satisfy onboard weapon effective range constraints. In the paper, the authors provide cost functions associated with degree of being in tail-chase to the target based on aspect and bearing angles as well as the cost functions associated with generating a maximum score when the opponent is within some threshold of the optimum shooting range. The cost functions are multiplied together to create the total cost. The cost functions are put into a receding horizon control scheme where the trajectory planning determined through selection of the controls  is performed for a given look-ahead time period where both players are utilising opposite strategies. Each player considers its opponents reachable sets within the horizon and uses this to select its choice of controls to maximise its payoff. This process is repeated every few control steps. For selecting the control actions, that authors mention the use the full set of control inputs within the performance envelope  rather than subset (e.g. turn, maintain hading, roll left at particular angle, immelman, split S or spiral dive) unlike most other works and thus point to generating a more optimal solution for each player selecting their respective strategies. Two simulation scenarios are provided with the first being the case where neither UAV starts off in air-superiority position and then exercises the receding horizon cost function optimisation to get into tail-chase with its opponent within optimum firing range. Neither succeed. The authors show the speed, load factor and bank-angle when applying the controls do not violate bounds during the flights and that feasible trajectories are generated. For the second simulation, the UAVs are initially in a tail-chase except not satisfying the within shooting range criterion. The opponent being chased manevouers to escape through applying the cost function while the chaser continues chasing. At the end of the engagement, within shooting range criteria are met and the target is directly in front but at the sub-optimal aspect, which leads to its escape. These scenarios are used to demonstrate the validity of the control strategy developed and thus provide the automatic selection of combat strategy for two unmanned aerial vehicles engaged in combat against one another. \\

Casbeer et al. \citep{casbeer2018target}, consider a scenario where an attacker missile pursuing an unmanned aerial vehicle target is engaged by two defending missiles launched from entities allied to the target which cooperate with the target. So, this scenario is an extension from a three party game where there is only one single defending missile engaging the attacker that cooperates with the target. The three party game is referred to as the Active Target Defence Differential Game (ATDDG). Besides computing the optimal strategy for the players in the extension to ATDDG, the paper is directed to determine the degree of reduction in vulnerability of the target when it uses two defenders rather than one. A constrained optimisation problem is formulated to setup the extension. It is shown that the target through having the choice to cooperate with either defender can more successfully escape the attacker. Additionally the presence of two defenders enables the attacker to be more easily intercepted. When the two defender missiles ae well positioned, both are able to intercept the attacker. \\

\subsection{Papers dealing with Cyber Warfare}

Keith et al. \citep{keith2021counterfactual} consider a multi-domain (cyber combined with air-defence) defence security game problem. There are two players engaging each other in a zero sum extensive form game, a defender, representing an integrated air-defence system combined with cyber security protection and an attacker, capable of unleashing air-to-ground threats (missiles, bombs) as well cyber-attacks (against IADS network). Here, the payoff has been selected as the expected loss of life. The defender wants to minimise this while attacker wants to maximise it. The cyber security game problem to protect the IADS is nested within the physical security game problem. The actions of the players correspond with allocations to activate IADS/cyber security responses nodes corresponding with population centres for the defender and allocations to attack IADS/associated cyber-security nodes by the attacker. The realism of the game is increased through provisioning in imperfect information, which is where the defender and attacker are not fully aware of vulnerability of nodes (selected randomly by a nature model which picks them in each game). Additionally, the defender is only able to sense cyber attacks on nodes probabilistically which has the implication that its allocation of cyber defence teams to particular IADS is only effective probabilistically. For the attacker, it can also determine the effectiveness of its cyber-attacks following physically attacking a node. This work is directed to advances security game literature by introducing the integrated domain and multiple periods for agent actions as well enabling continuous mixed form strategies. Also, the author considers it the first work where Monte Carlo (MC), discounted and robust counterfactural regret minimisation based approaches have been compared in security game that are applied and evaluated following increasing scale of the problem. Initially, for a small-scale problem the Nash equilibrium (NE) in the form of a sequence-form linear program is determined for the defender. Then, the problem is gradually scaled to expand to number of population centres to be defended up to an upper limit. Then the author, applies the approximate counterfactual regret minimisation (CFR) algorithm to reduce computation time while preserving the optimality of the strategy as much as possible. When the scale is further increased a discounted CFR is introduce which further reduces computation time. The parameter space of the problem and algorithm are explored to select best choice of tuning parameters and extract best performance per algorithm. The impact of limiting the rationality of the players, through introducing bounded rationality where the players do not necessarily make optimal robust best response moves. They can only manage approximate robust best response moves. A robust best response for a player is defined as the compromise between the completely conservative NE strategy and tje completely aggressive best response strategy.  It means that the strategy of the players has an exploitability where the opponent can capitalise on non-best responses being played by them. With respect to a player, the capability of their strategy to  capitalise over opponent is referred to as exploitation. Conversely, the vulnerability of their strategy with respect to an opponent is referred to as exploitability. When running all the different algorithms, the results show the Nash equilibrium solution being the safest strategy since the best moves are being played which are neither exploitable. The performance charts show that the robust linear program generates the highest mean utility and highest exploitation/exploitabiltiy ration while also introducing the maximum computational time. The Data biased CFR is seen to offer the best trade-ff by offering a high mean utility, a explotiation to exploitability ration in favour of exploitation while producing the lowest computational time.

\subsection{Papers dealing with Space Warfare}

In the domain of space warfare, human resources and risks are much less prevalent here, and instead the focus is more on network strength and interaction between independent autonomous agents, connected or otherwise. Ultimately, warfare in these aspects will operate at a pace and in dimensions far beyond human cognitive capacity. Since the rapidity and complexity of decisions within engagements will almost certainly outscale military personnel's understanding, game theory will take the place of decision makers as part of the overall software system, and imbue future technology to take human/ social factors into consideration when making calculations.  With a greater focus on connectivity and networking, the key to success in these areas relies on effective communication channels and a unanimously shared goal across the system. 

Zhong et al \citep{zhong2018joint} set the ambitious goal of optimising bandwidth allocation and transmission power across a satellite network. They base their research on bargaining game theory, and have to achieve compromise across: interference constraints, Quality of Service requirements, channel conditions, and transmission and reception capabilities for satellites at every point in the network. Interference constraints and bandwidth limitation are the surplus that need to be negotiated in the bargaining game, with each satellite using different strategies to improve their utility/ share of resources. This quickly escalates in complexity, with the most important takeaway from the model being the mapping of a problem to the cooperative bargaining game framework. 

Similarly, Qiao and Zhao \citep{qiaozhao} detail some key issues with the finite energy availability for the nodes in satellite networks. Their paper offers a solution through a game theoretical model of a routing algorithm, and use it to find an equilibrium solution to the uneven network flow. The model locates certain network hot spots which are reserving a lot of energy, and takes measures to evenly distribute the resources. This is another case of a bargaining/ cooperative game across multiple players in a network.

\subsection{Papers dealing with Target Tracking}


Gu et al. \citep{gu2010game} study the problem of tracking a moving target using a sensor network comprising of sensors capable of providing some position-related target measurements. Each sensor node has a sensor to observe the target and a processor to estimate its state. While there is some communication available among sensors, this ability is limited in the sense that each sensor node can only communicate with its neighbors. The problem is further compounded by the fact that the target is an intelligent agent capable of minimizing its detectability by the adversary and thereby has the potential to increase the tracking error of the tracking agent. Gu et al. \citep{gu2010game} solve this problem within the framework of a zero-sum game, and by minimizing the tracking agent's estimation error, a robust minimax filter is developed. Moreover, in order to handle the limited communication capability of the sensor nodes, they propose a distributed version of this filter for which each node only requires information in the form of current measurement and estimated state from its immediate neighbors. They then demonstrate the performance of their algorithm on a simulated scenario with an intelligent target and show that while the standard Kalman filter errors diverge, the minimax filter which takes into account the adversary's noise, is able to significantly outperform the Kalman filter.\\

Qilong et al.\citep{qilong2018} similarly address the issue of tracking an intelligent target, but they model a scenario where the tracking players are also in pursuit, and the focus is on protecting the target. Additionally, the target is able to fire a defensive missile at the attacker/ tracker. The attacker has line of sight of both the target and the defensive missile. The target's plan is to allow the tracker to slowly close the distance between itself and the target, all the while maneuvering to develop and understanding of how the attacker reacts. When the attacker is close to collision, the defensive missile is released. The target and the missile then  communicate, use the knowledge of the attacker's  movement patterns, and adhere to an optimal linear guidance law to destroy the attacker. This was modelled as a zero sum competitive game between the attacker, the target, and tjhe defensive missile. However, the paper also focuses on the cooperative game played between the target and the defensive missile, which is a non-zero sum game. For them, the payoff is calculated by minimised miss distance (which ideally equals zero - a collision with the attacker), as well as the control effort required to guide the defensive missile.

Faruqi \citep{faruqi2017differential} discusses the general problem of applying differential game theory to missile guidance. They state that missile trajectory follows Proportional Navigation (PN), a set of navigational systems using specific guidance laws for tracking targets. The performance of these systems is measured by a Linear System Quadratic Performance Index (LQPI). With respect to differential game theory, they model the  missile guidance problem by representing the missile navigation and trajectory with a set of differential equations. The general form of this problem is:
\begin{equation}
    \underline{\Dot{x}}_{ij} = F\underline{x}_{ij}(t) + G\underline{u}_{i}(t) - G\underline{u}_{j}(t)
\end{equation}
and
\begin{equation}
    J(...) = \frac{1}{2}\underline{x}_{ij}^{T}(t_{f})S\underline{x}_{ij}(t_{f}) + \frac{1}{2} \int_{t_{0}}{t_{f}} [\underline{x}_{ij}^{T}QG\underline{u}_{i} + \underline{u}_{i}^{T}R_{i}\underline{u}_{i} + \underline{u}_{j}^{T}R_{j}\underline{u}_{j}]
\end{equation}

where

\(\underline{x}_{ij}(t) = x_{i}(t) - x_{j}(t)\): is the relative state of player \(i\) w.r.t player \(j\)

\(\underline{u}_{i}(t)\): is the player \(i\) input

\(\underline{u}_{j}(t)\): is the player \(j\) input

\(F)\): is the state coefficient matrix

\(G\): is the player input coefficient matrix

\(Q)\): is the Performance Index (PI) weightings matrix for current relative states

\(S)\): is the PI weightings matrix for final relative states

\(R_{i}, R_{j}\): PI weightings matrices on inputs

Faruqi mainly focuses on two player and three player games, while the utility functions are modelled based on the relative distance vectors between missiles and targets. Faruqi shows that game theory can be effectively used in missile guidance tasks involving PNs in modern missiles.

Evers \citep{evers2013} on the other hand analyses defence against  Theater Ballistic Missiles (TBMs) using game theory. The proliferation of ballistic missiles and nuclear technology has important consequences for military conflict, where the cost of failure can lead to the destruction of entire cities. It can be hard to pinpoint their launch, since they have a big range are very powerful, though their payload can vary considerably. In combating this threat, the defending nation does have the advantage that there is usually  a long flight trajectory, typically split into three phases, during which the TBM can be intercepted. The Boost phase marks launch and the majority of the TBM ascent. The end of the boost phase is marked by the burn-out after which the TBM enters its midcourse phase. This phase is the longest phase of the flight, and affords the best opportunity for defenders to intercept the TBM. After the midcourse phase, the TBM enters its terminal phase from re-entry into the atmosphere. This is the last opportunity for defenders to intercept the missile. The flight path is illustrated in Figure  \ref{fig:flight} below.

\begin{figure*}[htbp]
\centering
\includegraphics[width=0.98\textwidth]{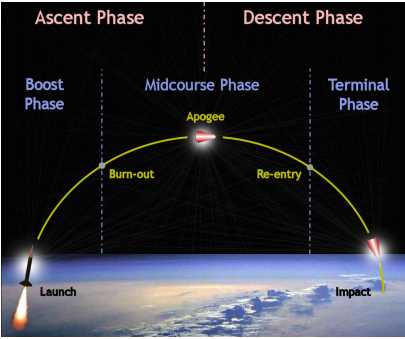}
\caption{Flight path of Theater Ballistic Missiles  \citep{evers2013}}
\label{fig:flight}
\end{figure*}

The missile travels a long distance over a reasonably extended flight time. However, from its physical geographical location, a defending military force or nation can only apply its resources to defend against it in the  termination phase of flight, where the risk is much higher and the cost of failure is at its greatest. For this reason, Evers proposes a cooperative strategy, where the defending nation forms alliances with nations around itself so that they too can attempt to intercept the TBM in its earlier phases as it travels to the impact location. Therefore, the game is divided into two smaller games: the first being a cooperative multiplayer game devising a set of strategies for the coalition of nations to utilise throughout a TBMs flight path, and the second is the bargaining and cooperation game between the defending nations and potential allies.

The basis of the cooperative game to shoot down the TBM is a strategy called 'shoot-look-shoot.' It relies on a set of \(N\) nations attacking the target using a set of strategies - their interceptor missiles - \(M\) each  of which has its own Probability of Interception \(P_{i}\). As the TBM flies, each nation  \(n\) in \(N\) will fire its missile(s) \(m_{n}\) to intercept the TBM, then look to see if it has successfully neutralised the threat. If it has failed, the next interceptor \(m_{n+1}\) will be fired. The game's problem is then reduced to optimising the probability of interception of the whole set of strategies, such that it has a feasible likelihood of stopping the TBM. Game theory is useful here because the principles of cooperative game theory provide a strong mathematical framework by which an equilibrium  solution can be reached for the set of cooperating nations as a whole. 

The second  game described is based on  negotiating with other nations to form an alliance. For these other nations, participation in this game is a risk because it makes  them another potential target for the attacking force. To solve this game, the defending nation must accurately evaluate the interceptor cost savings, that is, how much there is to gain by preventing the impact of the TBM. With these savings becoming the surplus that cooperating nations can share in, potential allies then negotiate over how those savings will be shared, in proportion to what interception resources they have to offer.

Shinar and Shima \citep{shinar1996game} continues the research of both pursuit-evasion games and ballistic missile defence with a zero sum game of a highly maneuverable ballistic missile avoiding an interceptor missile. More specifically, it ties in an imperfect information element to the game, where the ballistic missile knows it is being attacked by anti-missle missiles, but knows little about their  trajectory or launching location. In this game, the two players are the ballistic missile and the interceptor. If the ballistic missile uses a pure strategy, it will likely be hit because it either (a) cannot react quickly enough to an opponent it has little information about or (b) will move predictably and allow for a straightforward trajectory towards collision. Therefore, the best solution to the game for the ballistic missile is in mixed strategies. 

The mixed strategy will incorporate stochasticity in its flight pattern, assigning a probability distribution over a set of pure strategies. These pure strategies will be based on essential navigational heuristics, which will likely be known or easily discovered by the interceptor. By applying a small number of rapid and random switches in strategy, the ballistic missile  can maximise its potential for avoiding interception, and force the complexity of timing calculation back onto the interceptor.

Bogdanovic et al. \citep{bogdanovic2018target} investigate a target selection problem for multi-target tracking using a game-theoretic perspective. This is an important problem in a multi-function radar network as it needs to perform multiple functions such as volume surveillance and fire control simultaneously while effectively managing the available radar resources to achieve specified objectives. Thus, in effect they tackle a radar resource management problem in \citep{bogdanovic2018target} and use non-cooperative game-theoretic approaches to find optimal solutions for this problem. They formulate the problem in a framework where each radar is considered to be autonomous; there is no central control engine informing the radars of their optimal strategies nor is there any communication among the radars. First they consider a case where all radars share common interests with respect to the targets and for this problem they propose a distributed algorithm based on the best response dynamics to find the Nash equilibria points. This problem is then extended to a more realistic case of heterogeneous interests among radars and partial target observability. For this case, they employ the solution concept of correlated equilibria and propose an efficient distributed algorithm based on regret matching which is shown to achieve comparable performance to the more computationally intensive centralized approach.\\

Finally, Parras et al \citep{parras2017pursuit} examine a pursuit-evasion game, involving anti-jamming strategies for Unmanned Aerial Vehicles (UAVs). The game operates within a continuous time frame and is therefore dynamic, being solved with the help of differential Game Theory. In somewhat of a culmination of the aforementioned work, it combines elements of communication optimisation, sensor evasion and navigation. Given that UAVs require strong communication for control and the relaying of information, this dependency makes UAVs incredibly susceptible to jamming attacks. There are multiple strategies to both jam and anti-jam these communications, and this can be considered a zero-sum game where the UAV must attempt to optimise its communication capacity.  There is usually uncertainty about the  positioning and movement of the jamming agent, so the game is an imperfect information differential game. The most important payoff for the UAV is avoiding losing communication to jamming, and it  can do this by maneuvering to make approximations for the distance of jamming agents, and thus avoiding them.

\subsection{Papers dealing with Homeland Security}

The key components of homeland security addressed by game theory are cyber security, modelling terrorism threats and defence contracts. With many applications in computer science \citep{halpern2008computer, apt2011lectures, shoham2008computer}, game theory fits well into cyber security problems. Game theory combines the strict mathematical rigour of computer science, with more psychological and philosophical elements like attacker incentives and mindset, as well as human vulnerabilities in cyber security. Terrorism modelling similarly benefits the human considerations of game theory, since so much of their impact is not easily quantifiable, including social, economic and other spheres affected by terrorist threats, all of which are modled in a game-theoretic setting. These variables are necessary to understand the potential of terrorist threats and attacks, and to formulate strategies to best deal with them. Finally, game theory is suitable for a topic such as contracting and subcontracting because it captures interactions between selfish individuals effectively\citep{piraveenan2019applications, peldschus2008experience}.

The paper by Tom Litti \citep{tom2015game} provides a brief summary of how traditional network security heuristics can be updated with more precision, and how game theory can help network security engineers design strategies to properly predict, mitigate, and handle threatened networks. He develops a qualitative method for valuing the potential risks and costs of attacks on networks. While a fairly short paper, it does provide some cyber security situational examples of game theory in practice. For example, he models a two player zero sum game to represent an attacker and a security system. The nodes have their own interdependence, vulnerabilities and security assets, but cooperate to minimise the attacker's potential to compromise the system. This review is a great introduction to understanding how cyber security scenarios can be modeled in a game-theoretic framework.

Jhawar et al \citep{jhawar2016automating} cover a more specific approach of game theory, namely, Attack-Defence Trees (ADTs) to model scenarios. Here ADTs are used to map potential attack and defence scenarios on a system which is equipped with automatic defence protocols. The system needs to be comprehensive in addressing all possible vulnerabilities, as well as generating responses which adapt to the aggressively evolving situation of cyber security attacks. Currently, ADTs only provide upfront system analyses. Having a reactive strategy to cyber security is important because Attackers constantly change their attack strategies for offence, and so the split second judgements that are made in response can make the difference between a successful and a failed defence of a system.

In \citep{jhawar2016automating} they model a simple game of Attacker and Defender---a hacker and a security network administrator. The hacker attempts to breach the integrity of the system, and for each move they make, the administrator will have devised a reactive strategy based on the attacker's attempt. The greatest utility of this method comes from the ability to convert long extensive form games into a graphical layout for easier understanding and communication.

Gonzalez \citep{gonzalez2013individual} clearly outlines a standard two player competitive game of attacker and defender, and then addresses two aspects---Instance Based Learning Theory and Behavioural game theory. The former compiles cognitive information into a representation known as an instance. Each instance has a three part structure of situation, decision and utility---a standard game. However, it is the interaction between instances which is critical to this approach. Instance Based Learning Theory uses the learning from the outcome of each instance to feedback into the situation of the next instance, hopefully leading to better decisions in later iterations. This is notably similar to the reinforcement learning technique in machine learning. On the other hand, Behavioural game theory involves devising a strategy where we assess a variety of factors, to make more precise long term evaluations of targets and resources so that utility scores more closely reflect real life value. Once again, game theory allows us to draw upon social information in cyber security situations, and assess how that will affect behaviour for both players in the game. Other such factors include motivational factors of the players, completeness of information for each player, technological constraints and inefficiencies between player and technology. Gonzalez stresses that the importance of accommodating for these factors in any cyber security model will help make more realistic and useful policies for cyber defence.

One common use of cyber security is in the prevention of terrorism. Hausken et al \citep{Hausken2009} cover both terrorism and natural disaster modelling with some guiding game theory principles. Terrorism and Natural Disasters are addressed by defending with anti-terrorism, anti-disaster, and anti-all-hazard investments. Making projections about the likelihoods of each of these occurrences, defenders must make strategic decisions towards the amount of investment dedicated to each defence. Costs to be considered in the utility function of each of these situations include: the intelligence of terrorists or the randomness/ environmental control of natural disasters; the intensity of an attack/ disaster; and differences in evaluation of target value between a terrorist and defender. The game-theoretic approach used in this analysis captures the defender's effort in combating each threat. Depending on the likelihood of each event, combined with the costs of each defensive system, defenders can derive an optimal division of funds.

Kanturska et al. \citep{Kanturska2009} write a rigorous inspection of how transport network reliability can be assessed using game theory, when attack probability on different locations is unknown. The approach favoured using a minimax algorithm to distribute risk across multiple paths as long as the travel costs are small relative to potential losses incurred by an attack. This would be useful in negating the potential risk associated with safely escorting a VIP through a city. Game theory is helpful in this situation because it can analyse network reliability when the attack probabilities are unknown.

Bier \citep{Bier2006} presents useful game theory based suggestions for policy insights and investment decisions, insurance policy premiums and more. Her work discusses the weakest link model: a strategy focusing all resources on preventing the worst utility scenario. This is generally weak in practice, and she suggests instead to consider hedging those investments with a variety of defense strategies for different potential targets. The paper looks into standard terrorist/ defender games, and how security investments can change the landscape of attacker-defender interactions for the whole community. This is mainly done through its own scoping study, with one of the key takeaways being that terrorism mitigation systems can benefit from game theory because it adds an extra layer of consideration of the terrorists' response to any defence mechanisms. Hence, game theory, combined with the holistic approach of risk and reliability analysis over all systems, can provide a more comprehensive assessment of all the potential risks and vulnerabilities in counter terrorism strategies.

Cioaca \citep{cioacua2013considerations} investigates a similar question to Bier et al \citep{bier2008} as mentioned earlier, but specifically towards aviation security. For Cioaca, the problem is summarised by targeting both the cost of security measures for airports and the cost of maintaining a stable and resilient system of defence. Key strategies are: preventing the attack or threat entirely (perhaps by removing all access to the target location or restricting airlines permissions if they fail to adhere to guidelines to helping); managing the temporal dimension of the attack (the length of the attack and the subsequent time to recover from it); understanding all direct and indirect losses (both casualties and related damages like contamination or infection, compromised secondary security measures, or reputational/ signal ramifications); and the costs of mitigation, response and recovery. 

The model is built around several factors and parameters. The first and most critical is human losses and material damages \(C(H,D|b_{t})\). The most obvious and direct casualties of an attack, these two losses are highly negative payoffs in such an attack, and are often greater than any cost to prevent them. Human losses \(H\) are hard to numerically quantify, and therefore when making proper assessments about resource division, understanding how to minimise human losses across different groups of humans and in different dimensions is one of the most difficult aspects of this problem. Material damages \(D\) can be monetarily quantified, but often the run-on effects of such damages are where the significant losses are incurred. These losses can lead to total infrastructure shutdowns,  ceasing operation of the facility, loss of jobs of workers and even potentially the slow decline and shutdown of the facility altogether. The second major factor in this game is the budget allocated to security systems \(b_{t}\). Organisational and managing bodies will only have a certain amount of resources allocated to a security system \(T\).  The next factor is the number \(s\) of security system components as this will be how the budget is composed. Each of these components are partitioned into one of \(n\) separate system sub-components. These components are divided amongst a number of targets \(t_{ik}\), and each one of these targets is assigned a probability of being attacked \(p_{a_{j}}^{t_{k}}(b_{t_{k}})\) and a value \(w_{t_{ik}}\). This can be formally expressed as:
\begin{equation}
    \min C(H,D|b_{t}) = \sum_{j=1}^{l}  \sum_{i=1}^{n}  \sum_{k=1}^{s} p_{a_{j}}^{t_{ik}} \cdot p_{v_{j}}^{t_{ik}} \cdot w_{t_{ik}}
\end{equation}

For any system of resource division, Ciaoca advocates establishing dimensions of measuring system resilience. This is divided into static resilience, the efficient allocation of resources; and dynamic resilience, recovery speed of the system after the shock, including long term investment in-flows. These two forms of resilience signify the strength of a system both before, during and after an attack. With respect to game theory, Ciaoca's study defines a game clearly, and  incorporates a myriad of complex and interconnected parameters to outline an effective and calculable game model.

The final paper we discuss on home land security  is by Gardener and Moffat \citep{gardener2008}.  A distinct and unique approach to game theory in defence, this paper covers the notion of developing a strategy to assess defence contractors and their potential performance/ ability to meet contractual obligations. In game theory terms, this problem is a question of cooperation and defection. It suggests quantitative methods through which defence departments can more rigorously assess contracts and bidding scenarios, and therefore wisely select contractors, and protect their budget. Gardener and Moffat further the understanding of change requirements in project management at different bidding stages of defence acquisition projects. The factor they focus on is the \textit{conspiracy of optimism}, whereby projects spiral out of control---past budget limits and necessary deadlines---due to irrational expectations of project progress. Often this 'conspiracy' is a drive towards making short term gains, and in fact leads to overall losses. The bidding game that is played becomes less about the success of the project, and more about profit capitalisation. The paper presents a guided approach to understanding how 'cooperative' games still lead to selfish motivations. Such is the unreliability of contract outcomes that the game further degenerates to a two player game against the industry of contractors as a whole.

\subsection{Papers dealing with Other /  Mixed warfare}

Zhang and Meherjerdi \citep{zhang2013survey} investigate how groups of multiple unmanned vehicles can be used and controlled using game-theoretic methods in different communication frameworks. Dividing a mission for a single unmanned vehicle amongst multiple unmanned vehicles yields more effective task allocation and performance. The separation of labour away from one powerful single vehicle to several smaller vehicles provides flexibility, adaptability and improved fault tolerance. The uses of such a network are surveillance, exploration, satellite clustering, combining Unmanned Underwater Vehicles (UUVs) and submarines, planes and UAVs, and cooperative robot reconnaissance. As evident from this list, the strategy is incredibly powerful because if its ability to combine resources across multiple domains. To be able to operate a cohesive unit of resources across multiple unmanned systems for combat or exploration brings an unprecedented level of information and control which would accelerate military performance tremendously.

Search is a `hide and seek' game with a long history in military applications \citep{stone1981search,   stone1980optimal, washburn2014two,  washburn2015multistatic, jing2021recent,  ristic2020intermittent, ristic2016study, ristic2020decentralised, hutchinson2018information, park2020cooperative}. The theory was pioneered by Koopman \citep{stone1981search} primarily in the military context (search for an escaping target), followed by developments by Stone \textit{at al} \citep{stone1980optimal}. The applications include  submarine hunting, mine detection, rescue operations, risk for the first responders, and backtracking of a hazardous source \citep{stone1981search,   stone1980optimal, washburn2014two,  washburn2015multistatic, jing2021recent,  ristic2020intermittent, ristic2016study, ristic2020decentralised, hutchinson2018information, park2020cooperative}.  This framework provides the optimal  \textit{a priori} search plan for a given detection model, target motion and the cost of the search. The cost of the search may include time of the search, probability of escape (for a target), exposure risk (for a searcher), information entropy, or situation awareness (map of probability of target location). The searcher can be a moving platforms (UAV, UV, patrolling boats, helicopters, robots, people) and the targets can be static, movable, blind, silent, or  emitting. In this context the simultaneous localisation and mapping (SLAM) algorithms are often used \citep{durrant2006simultaneous}.      

The new direction of this research (inspired by some biological applications) employs the ideas of infotaxis \citep{vergassola2007infotaxis}, or real-time control of the searchers movement based on information (entropy) gain  extracted from the environment (sporadic measurements, forbidden areas, communication between searchers).  This method resembles   analogy with navigation of some animals \citep{vergassola2007infotaxis,benichou2011intermittent}. \\

\section{Classification of papers} \label{classification}

In the previous section, it would have been noticeable that many papers have applicability in more than one domain, and use myriad types of games and model a range of players. It is therefore imperative to classify the reviewed papers in a principled manner. To do so, we use the classification scheme already introduced in Table \ref{classstructure} in section \ref{background}. 

In particular, the reviewed papers could be classified based on (1) the domain or type of warfare (2) the type of game or games used in the paper (3) the nature of players modelled in the paper. The domain can be broadly classified into Traditional (T) or Modern (M), and more specifically into Land warfare,  Naval  warfare, Aerial  warfare, Cyber warfare and Space Warfare. The type of games used can also have a complex classification, based on whether the games were Non-cooperative or cooperative, sequential or simultaneous,  discrete or continuous,  zero sum or non-zero sum. Finally, the games can be two-player, three-player, or multi-player games. All of this is succinctly captured in Table \ref{classstructure}.

In Table \ref{table:MilGTPapersClassified}, we provide a self-explanatory, elaborate classification of all the reviewed papers based on the above-mentioned classification scheme.


\begin{longtable}[c]{|l|c|l|}
\caption{Defence Game Theory Classifed using Table~\ref{classstructure} }
\label{table:MilGTPapersClassified}\\
\hline
\multicolumn{1}{|c|}{\textbf{Title}} & \textbf{Authors} & \multicolumn{1}{c|}{\textbf{\begin{tabular}[c]{@{}c@{}}Classification Code using\\ Military Defence \\ Game Theory\\ Classification System\end{tabular}}} \\ \hline
\endfirsthead
\endhead
\multicolumn{1}{|c|}{Game Theoretic analysis of adaptive radar jamming} & Bachmann et al & \multicolumn{1}{c|}{T-N-IW-NCo-Sim-D-ZS-} \\ \hline
\begin{tabular}[c]{@{}l@{}}Target selection for tracking in multifunction radar\\ networks: Nash and Correlated equilibria\end{tabular} & Bogdanovic et al & T-A-IW-NCo-Sim-D-ZS- \\ \hline
\begin{tabular}[c]{@{}l@{}}Power allocation game between a radar network\\ and multiple jammers\end{tabular} & Deligiannis et al & T-A-IW-NCo-Sim-D-ZS- \\ \hline
\begin{tabular}[c]{@{}l@{}}Strategies for defending a coastline against multiple\\ attackers\end{tabular} & Garcia et al & T-A-IW-NCo-Sim-D-ZS- \\ \hline
\begin{tabular}[c]{@{}l@{}}A game theory approach to target tracking in\\ sensor networks\end{tabular} & Gu et al & T-L-IW-NCo-Sim-D-ZS- \\ \hline
\begin{tabular}[c]{@{}l@{}}Joint Power allocation and beamforming between \\ a multistaticradar and jammer based on game theory\end{tabular} & He et al & T-A-IW-NCo-Sim-D-ZS- \\ \hline
\begin{tabular}[c]{@{}l@{}}Game theoretic situation and transmission in \\ unattended ground sensor networks: a correlated\\ equilibrium approach\end{tabular} & Krishnamurthy et al & T-L-(RAW \& AMW)-NCo-Sim-D-ZS- \\ \hline
\begin{tabular}[c]{@{}l@{}}Network enabled missile deflection: games and \\ correlated equilibrium\end{tabular} & Maskery et al & T-N-(RAW \& IW)-Co-Sim-D-NZS- \\ \hline
\begin{tabular}[c]{@{}l@{}}Decentralised algorithms for netcentric Force \\ Protection against anti-ship missiles\end{tabular} & Maskery et al & T--N-(RAW \& IW)-NCo-Sim-D-ZS- \\ \hline
Stochastic game approach to air operation & McKeaney et al & T-A-RAW-NCo-Sim-D-ZS- \\ \hline
\begin{tabular}[c]{@{}l@{}}Hybrid Game Theory and D-S Evidence Approach\\ to Multiple UCAVs Cooperative Air Combat \\ Decision\end{tabular} & Xingxing Wie et al & T-A-RAW-NCo-Sim-D-ZS- \\ \hline
\begin{tabular}[c]{@{}l@{}}Cooperative Occupancy Decision Making of \\ Multiple-UAV in Beyond-Visual Range Air-Combat:\\ A Game Theory Approach\end{tabular} & Yingying Ma et al & T-A-RAW-NCo-Slt-D-ZS- \\ \hline
\begin{tabular}[c]{@{}l@{}}A Simple Game-Theoretic Approach to Suppression\\ of Enemy Defenses and Other Time Critical Target\\ Analyses\end{tabular} & \multicolumn{1}{l|}{Thomas Hamilton et al} & T-A-RAW-NCo-Slt-D-ZS- \\ \hline
The Cooperative Ballistic Missile Defence Game & Lanah Evers et al & T-A-RAW-NCo-Slt-D-ZS- \\ \hline
\begin{tabular}[c]{@{}l@{}}Cooperative Missile Guidance for Active Defense \\ of Air Vehicles\end{tabular} & Eloy Garcia et al & T-A-RAW-NCo-Slt-D-ZS- \\ \hline
\begin{tabular}[c]{@{}l@{}}Counterfactual regret minimization for integrated \\ cyber and air defense resource allocation\end{tabular} & Andrew Keith et al & M-C-RAW-NCo-Slt-D-ZS- \\ \hline
\begin{tabular}[c]{@{}l@{}}Game Theoretic Model for the Optimal Disposition \\ of Integrated Air Defense System Assets\end{tabular} & Chan Y. Han et al & T-A-RAW-NCo-Seq-D-ZS- \\ \hline
\begin{tabular}[c]{@{}l@{}}Differential game theory and applications \\ to Missile Guidance\end{tabular} & Faruqi & T-A-WAW-NCo-Slt-C-ZS-(2P,3P,NP) \\ \hline
\begin{tabular}[c]{@{}l@{}}Assessment of Aerial Combat Game via \\ Optimisation-Based Horizon Control\end{tabular} & Baspınar, Barıs et al & T-A-RAW-NCo-Slt-D-ZS- \\ \hline
Target Differential Game with Two Defenders & David Casbeer & T-A-WAW-NCo-Slt-C-ZS-3P \\ \hline
The optimal search for weak targets & \multicolumn{1}{l|}{Koopman, Stone et al} & T-N-AMW-NCo-Slt-D-ZS- \\ \hline
\begin{tabular}[c]{@{}l@{}}Optimal Strategy for Target Protection with a \\ defender in the pursuit-evasion scenario\end{tabular} & Qilong et al &  \\ \hline
\begin{tabular}[c]{@{}l@{}}Differential game theory with applications to\\  missiles and autonomous systems guidance\end{tabular} & Faruqi & T-A-(AMW,WCW)-NCo-Slt-C-ZS- \\ \hline
\begin{tabular}[c]{@{}l@{}}A game theoretical interceptor guidance law \\ for ballistic missile defence.\end{tabular} & Shinar et al & T-A-(AMW,WCW)-NCo-Slt-C-ZS- \\ \hline
\begin{tabular}[c]{@{}l@{}}Pursuit-Evasion games: a tractable framework for\\ anti-jamming in aerial attacks\end{tabular} & Parras et al & T-A-AMW-Nco-Slt-C-ZS- \\ \hline
\begin{tabular}[c]{@{}l@{}}A simple game theoretic approach to suppression of\\ enemy defences and other time-critical target\\ analyses\end{tabular} & Hamilton et al & T-A-(RAW,WCW)-NCo-Slt-D-ZS- \\ \hline
\begin{tabular}[c]{@{}l@{}}Choosing What to Protect: Strategic Defence\\ allocation against an unknown attacker\end{tabular} & Bier et al & T-L-RAW-NCo-Slt-D-ZS- \\ \hline
\begin{tabular}[c]{@{}l@{}}Considerations on Optimal Resource allocation\\ in avation security\end{tabular} & Cioaca & T-L-RAW-NCo-Slt-D-ZS- \\ \hline
\begin{tabular}[c]{@{}l@{}}Horsemen and the empty city: A game theoretic\\ examination of deception in Chinese military\\ legend\end{tabular} & Cotten et al & T-L-IW-NCo-Slt-D-ZS- \\ \hline
An Economic Theory of Destabilisation War & Gries et al & T-L-IW-NCo-Seq-D-ZS- \\ \hline
\begin{tabular}[c]{@{}l@{}}Game theoretic approach towards network security:\\ A review\end{tabular} & Tom Litti & M-C-IW-NCo-Slt-D-ZS- \\ \hline
\begin{tabular}[c]{@{}l@{}}Automating cyber defence responses using \\ attack-defence trees and game theory\end{tabular} & Jhawar et al & M-C-WCW-NCo-Slt-D-ZS- \\ \hline
\begin{tabular}[c]{@{}l@{}}From individual decisions from experience to\\ behavioural game theory\end{tabular} & Gonzalez & M-C-IW-NCo-Slt-D-ZS- \\ \hline
\begin{tabular}[c]{@{}l@{}}Game Theoretic Approaches to Attack Surface \\ Shifting. Moving Target Defense II\end{tabular} & Manadhata &  \\ \hline
\begin{tabular}[c]{@{}l@{}}Improving reliability through Multi-Path routing\\ and Link Defence: An Application of Game Theory\\ to Transport\end{tabular} & Kanturska et al & M-C-WCW-NCo-Slt-D-ZS- \\ \hline
\begin{tabular}[c]{@{}l@{}}Game Theoretic and Reliability models in \\ counter-terrorism and security\end{tabular} & Bier et al & M-C-(IW,AMW,WCW)-NCo-Slt-D-ZS- \\ \hline
\begin{tabular}[c]{@{}l@{}}Changing behaviours in defence acqusition: \\ a game theory approach\end{tabular} & Gardener et al &  \\ \hline
\begin{tabular}[c]{@{}l@{}}Joint Transmit Power and Bandwidth Allocation \\ for Cognitive Satellite Network based on Bargaining\\ Game Theory\end{tabular} & Zhong et al & M-S-RAW-Co-Slt-D-NZS \\ \hline
\begin{tabular}[c]{@{}l@{}}The Research and Simulation of a Satellite Routing\\ Algorithm based on Game Theory\end{tabular} & Qiao et al & M-S-WCW-NCo-Slt-D-ZS \\ \hline
\begin{tabular}[c]{@{}l@{}}A survey of multiple unmanned vehicles formation\\ control and coordiation. Normal and fault situations\end{tabular} & Zhang et al & T-A-WCW-Co-Slt-D-NZS \\ \hline
\end{longtable}

\section{Opportunities for Further Research} \label{futurework}

The reviewed papers haver shown that game theory can provide a unifying framework to analyse decision making behaviours of agents in defence contexts.  In this section, we briefly discuss a range of potential defence scenarios where game theory has hitherto not been applied, but would make a useful contribution if applied in future.

A recent investigation by Defence Advanced Research Projects Agency (DARPA)  into `Mosaic warfare'  \citep{Darpa} is an example of such a potential future application of game theory.  The idea is mentioned in Zhang \citep{zhang2013survey} in the context of operating multiple unmanned vehicles, and proposes having a lot of smaller cost-efficient resources interconnected in a 'mosaic' network, such that if several units are destroyed, the overall nature of the network stands, much like how a mosaic retains its image even if a few tiles are removed.  The goal is that such a vast array of resources with different capabilities will be able to overwhelm the enemy with its completeness and complexity. It utilises principles of concurrency to address the intricacy of the connections in  systems of millions of sensors and actuators.  These systems in turn must handle inter-system communication.  If successfully implemented, such a system of systems can provide military strategists with an overwhelmingly powerful network of weaponry and resources,  which can defeat opponents with the sheer scale and complexity of their dynamics.  This method of combining different parts of the arsenal maximises the benefits of each, and reintroduces a focus on expendability of resources, rather than focussing on a few pieces of high value weaponry.  This  in turn builds resilience and adaptability into the strategy, a shift away from heavyweight, single focus attack methods. Since there are a high number of lower cost resources which need to cooperate for the best outcome, this scenario could be modelled as a multi-player co-operative game at one level, while the strife with the opponent(s) can be modelled as a muti-player non-cooperative game.  it cane be noted that the concept of `Mosaic warfare' is essentially similar to the more general concept of agent-based modelling, which has already been used in several diverse contexts, ranging from Ageless Aerospace Vehicle design  \citep{prokopenko2005convergence, prokopenko2005towards} to modelling of infectious disease dynamics \citep{cliff2018investigating}, and game theory has already been used successfully in some of these contexts \citep{chang2020game, piraveenan2021optimal}.

Another area, within  the context of naval warfare, where game theory could be fruitfully applied in naval susceptibility. in analysing naval susceptibility,  naval vessels factor in their environment, movement patterns and potential adversary sensors to calculate their risk of detection when moving covertly \citep{suscep2010}.  Such an application has overlaps between commonly studied tracking problems in defence science,  as explained in  Gu's \citep{gu2010game}    which describes tracking using sensor networks.  Such a scenario, as elaborated before, could be modelled as a two-player non-cooperative differential game, with detection being the main payoff parameter for each player.

Indeed, ground based tracking problems could benefit from the application of  game theory as well, and papers in this area so far has been few. Ground based tracking problems may appear in both ground based  military applications (classified here as ground warfare), as well as domestic security and anti-terrorism applications (classified here as home-front warfare),  where the  the ability  of security agencies to track individuals'  movements throughout a society - including their locations, their social network and their motivations - is a crucial capability\citep{zalud2010}. The later scenario could be modelled as a two player game of pursuit and evasion, or perhaps just pursuit and reconnaissance with the aim of not revealing the pursuit to the target, while the target would try to identify pursuit.  The amount of predictive information gained from covert tracking would be the  payoff in this situation. 

Modelling cyber warfare is another area where game theory could be applicable, and again there have been few papers addressing this niche, other than papers coming primarily from the computer science domain and focus primarily on cyber security.  Kim et al \citep{kim2019leading} describe cyber warfare scenarios which are integral to all military operations, and highlight the critical role played by new technological paradigms such as Internet of Things (IoT) and Brain Computer Interfaces. Defence experts increasingly have a need to predict and preempt cyber warfare strategies of hostile players. Modelling decision making scenarios involving  Cyber warfare scenarios with novel technological interfaces is an area where game theory can play a vital role.

\section{Conclusions} \label{conclusions}

Game theory has proven itself as a versatile and powerful tool for obtaining vital insights  about the decision making processes of agents and players in a number of  fields. In this review paper, we elaborated several scenarios in which game theory could be applied in defence science and technology, and presented a succinct review of existing research in this direction.  We introduced an extensive classification for the reviewed papers, based on the kind of warfare studied, type of games used, and the nature of players.  Based on the observations made, we identified gaps in the literature where game theory has not been applied extensively so far but has great potential to be applied fruitfully in future, and we discussed the potential directions in which defence applications of game theory could expand in the future

The domain based classification was the primary mode of classification that employed, and in this context we grouped the reviewed papers into papers about  tracking systems, aerial warfare, ground warfare, home front warfare and space/ other warfare.  For each paper considered, the number and roles of players and game types were defined, and where possible, strategies and payoff functions were discussed. The goal of this exercise was to identify the most commonly analysed domains as well as frequently used game types, and use this knowledge to identify gaps in literature and cross-pollinate ideas across various domains and types of warfare within the defence context.

As the world deals with emerging challenges to peace and stability, the future of humanity depends on our ability to solve problems peacefully. While this is a lofty goal to achieve, the projection of power is  decidedly better than an actual armed conflict which will be very costly at many levels, and game theory could indeed play a part in deciding some of the `soft conflicts' which could play out in the coming years and decades. As the focus on defence strategies and capabilities are likely to increase in the coming years, game theory can serve as an additional tool that defence scientists can use at many levels of abstraction to solve deployment, sensing, tracking, and resource allocation problems.



\authorcontributions{MP, SA, and AS conceptualised the research. All authors undertook the literature survey  and wrote the manuscript.}


\acknowledgments{ The authors thank Dr. Farhan Faruqui from the Defence Science and Technology Group, Australia for many fruitful discussions}

\conflictsofinterest{The authors declare no conflict of interest.} 

%

%
%
\reftitle{References}



\externalbibliography{yes}
\bibliography{game_defence_review}





\end{document}